\documentclass[twocolumn,aps,prl,showpacs,amsmath,amssymb,floatfix]{revtex4-2}

\usepackage{filecontents}
\usepackage[T1]{fontenc}
\usepackage[utf8]{inputenc}
\usepackage[english]{babel}
\usepackage{dsfont}
\usepackage{relsize}
\usepackage{xcolor}
\usepackage{float}
\usepackage{float}
\usepackage{graphicx}
\usepackage{dcolumn}
\usepackage{bm}
\usepackage{xcite}
\usepackage{xr}
\makeatletter
\newcommand*{\addFileDependency}[1]{ 
  \typeout{(#1)}
  \@addtofilelist{#1}
  \IfFileExists{#1}{}{\typeout{No file #1.}}
}
\makeatother

\newcommand*{\myexternaldocument}[1]{%
    \externaldocument[SM-]{#1}%
    \addFileDependency{#1.tex}%
    \addFileDependency{#1.aux}%
}
\myexternaldocument{./supplementary_material_v2}

\newcommand{\quotes}[1]{``#1''}

\renewcommand{\figurename}{FIG.}
\addto\captionsenglish{\renewcommand{\figurename}{FIG.}}

\usepackage{subfigure}

\usepackage{comment}

\usepackage[colorlinks]{hyperref}
\hypersetup{
	colorlinks = true,
	linkcolor = red,
	citecolor = {green},
	urlcolor = {blue}
}

\begin{document}


\title{Collective excitations of a strongly-correlated non-equilibrium photon fluid \texorpdfstring{\\}{} across the Mott/superfluid phase transition
}

\author{Fabio Caleffi$^1$}
\email{bafioc11@gmail.com}
\author{Massimo Capone$^{1, 2}$}
\author{Iacopo Carusotto$^{3}$}
\affiliation{$^1$International School for Advanced Studies (SISSA), Via Bonomea 265, I-34136 Trieste, Italy}
\affiliation{$^2$CNR-IOM Democritos, Via Bonomea 265, I-34136 Trieste, Italy}
\affiliation{$^3$INO-CNR BEC Center and Dipartimento di Fisica, Universit\`a di Trento, Via Sommarive 14, I-38123 Povo, Italy}

\date{\today}


\begin{abstract}
We develop a Gutzwiller theory 
for the non-equilibrium steady states of a strongly-interacting photon fluid driven by a non-Markovian incoherent pump.
In particular, we explore the collective excitation modes across the out-of-equilibrium Mott/superfluid transition, characterizing the diffusive Goldstone mode in the superfluid phase and the particle/hole excitations in the insulating one. Observable features in the pump-and-probe optical response of the system are highlighted. Our results appear as experimentally accessible to state-of-the-art circuit-QED devices and open the way for driven-dissipative fluids of light as quantum simulators of novel many-body scenarios.
\end{abstract}

\maketitle









\textit{Introduction.}
Quantum fluids of light with effective photon interactions are rapidly growing as a new branch of many-body physics
~\cite{RevModPhys.85.299, hartmann}. Right after the observation of Bose-Einstein condensation~\cite{science.1140990}, superfluidity~\cite{Amo2009} and hydrodynamic generation of topological excitations~\cite{Nardin2011, Sanvitto2011} in polariton fluids in semiconductor microcavities, exciting advances in circuit-QED engineering~\cite{PhysRevLett.79.1467, PhysRevLett.101.246809, Fink2008} are preparing the ground for the exploration of strongly-interacting fluids~\cite{Houck2012, google_1, google_3, Carusotto2020} and, consequently, the quantum simulation of bosonic lattice models~\cite{Hartmann2006, Greentree2006, PhysRevB.90.195112, noh}. Specifically, these systems
appear as a new platform to explore Bose-Hubbard physics~\cite{fisher} and the Mott/superfluid quantum phase transition~\cite{PhysRevA.76.031805, PhysRevA.81.061801, PhysRevA.84.043827, You:19} in a novel out-of-equilibrium context. 

In this regard, pioneering theoretical investigations have explored the rich variety of non-equilibrium steady states (NESS) under coherent or Markovian pumping protocols~\cite{PhysRevLett.99.186401, PhysRevLett.103.033601, PhysRevLett.108.233603, PhysRevLett.110.233601}, also in comparison with the corresponding equilibrium systems~\cite{PhysRevA.90.063821}. 
Non-Markovian scenarios~\cite{kapit2014induced, PhysRevA.96.023839, PhysRevA.96.033828} provide us with a whole new direction~\cite{PhysRevResearch.2.043232}, as they proved crucial for the experimental creation of genuine Mott insulating states~\cite{Ma2019}.

These experimental advances call for theoretical approaches able to investigate the nature of the observed steady states. This is a challenging task bridging the quantum optics, condensed matter and many-body communities, which therefore requires to establish a common language and an interdisciplinary perspective. Among the open issues, we mention the collective properties and dynamical correlations of out-of-equilibrium insulators and strongly-interacting superfluids. Even though various techniques are available to describe driven-dissipative systems, such as variational methods~\cite{PhysRevB.99.214306, PhysRevLett.114.040402, PhysRevA.96.023839, PhysRevLett.122.250502, PhysRevLett.122.250503}, matrix-product states~\cite{PhysRevLett.93.207204, PhysRevLett.93.207205, PhysRevA.96.033828, PhysRevLett.122.043602, Kshetrimayum2017, PhysRevLett.124.043601} and clustering techniques~\cite{PhysRevX.6.031011}, major hurdles still persist in the extension of these methods to the dynamics of large and high-dimensional non-Markovian systems~\cite{PhysRevA.96.023839, PhysRevA.96.033828, PhysRevResearch.2.043232}.


In this Letter, we make use of the Gutzwiller ansatz -- a powerful description of Mott insulator states in generic condensed matter systems~\cite{PhysRev.137.A1726, PhysRevB.2.4302, PhysRevB.44.10328, PhysRevB.45.3137} -- and we extend it to the non-equilibrium context of strongly-interacting photons in a cavity array under a non-Markovian incoherent pump. The evolution of the collective excitation spectrum across the insulating and superfluid phases is characterized, and novel features
stemming from the non-equilibrium condition are highlighted.
Our predictions represent a first step towards the understanding of quantum fluctuations in strongly-correlated out-of-equilibrium many-body systems. Observable fingerprints are identified in the response of the system to additional weak probes, a quantity directly accessible to experiments with state-of-the-art circuit-QED technology.


\textit{Model and mean-field theory.}
We consider a $d$-dimensional array of coupled optical cavities modeled by a Bose-Hubbard (BH) Hamiltonian,
\begin{equation}\label{eq: BH_model}
    \hat{H}_{\mathrm{BH}} = \sum_{\mathbf{r}} \left( \omega_c \, \hat{a}^{\dagger}_{\mathbf{r}} \, \hat{a}_{\mathbf{r}} + U \, \hat{a}^{\dagger}_{\mathbf{r}} \, \hat{a}^{\dagger}_{\mathbf{r}} \, \hat{a}_{\mathbf{r}} \, \hat{a}_{\mathbf{r}} \right) - J \sum_{\langle \mathbf{r}, \mathbf{s} \rangle} \hat{a}^{\dagger}_{\mathbf{r}} \, \hat{a}_{\mathbf{s}} \, ,
\end{equation}
where $\hat{a}_{\mathbf{r}} \left( \hat{a}^{\dagger}_{\mathbf{r}} \right)$ is the annihilation (creation) operator associated with the cavity mode at site $\mathbf{r}$, $J$ is the hopping energy, $\omega_c$ is the bare cavity frequency and $U$ is the photon-photon interaction energy stemming from the optical non-linearity of the cavity medium~\cite{RevModPhys.85.299, hartmann, Carusotto2020}.

The driven-dissipative dynamics of the BH array is ruled by photon losses at a rate $\Gamma_l$ and the coupling of each cavity mode to incoherently pumped two-level emitters (TLE's). The dynamics of the TLE's coupled to the cavities is governed by the Hamiltonian
\begin{equation}\label{eq: H_emitters}
    \hat{H}_{\mathrm{em}} = \omega_{\mathrm{at}} \sum_{\mathbf{r}} \hat{\sigma}^+_{\mathbf{r}} \, \hat{\sigma}^-_{\mathbf{r}} + \Omega \sum_{\mathbf{r}} \left( \hat{a}^{\dagger}_{\mathbf{r}} \, \hat{\sigma}^-_{\mathbf{r}} + \text{h.c.} \right) \, ,
\end{equation}
where $\hat{\sigma}^{\pm}_{\mathbf{r}}$ is the rising (lowering) operator in the pseudospin space of each TLE. These are pumped via some Markovian mechanism at a rate $\Gamma_p$ and decay at a rate $\gamma$. In order to achieve an efficient pumping mechanism for the cavity modes, we assume $\Gamma_p \gg \gamma$, which results in population-inverted TLE's. Most notably, this provides a straightforward realization of a non-Markovian driving protocol for the cavity modes, with an energy-dependent gain~\cite{PhysRevA.96.023839} which is at the roots of various non-equilibrium quantum critical regimes~\cite{PhysRevLett.116.070407, Sieberer_2016} and has been predicted to give a high-fidelity emulation of the BH model~\cite{PhysRevA.96.033828}.

\begin{figure}
    \centering
    \includegraphics[width=0.75\linewidth]{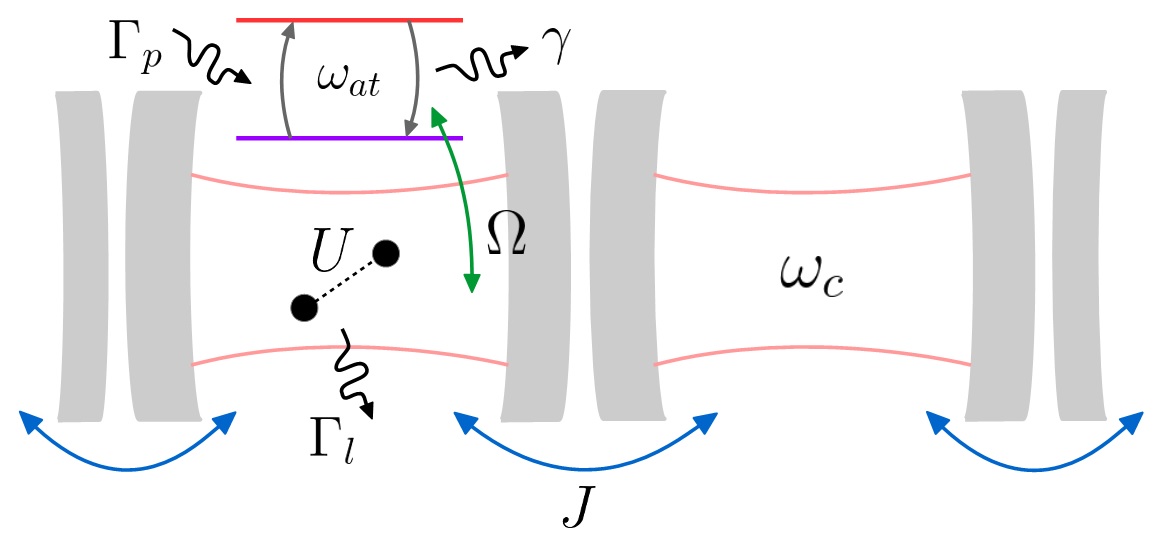}
    \includegraphics[width=1.0\linewidth]{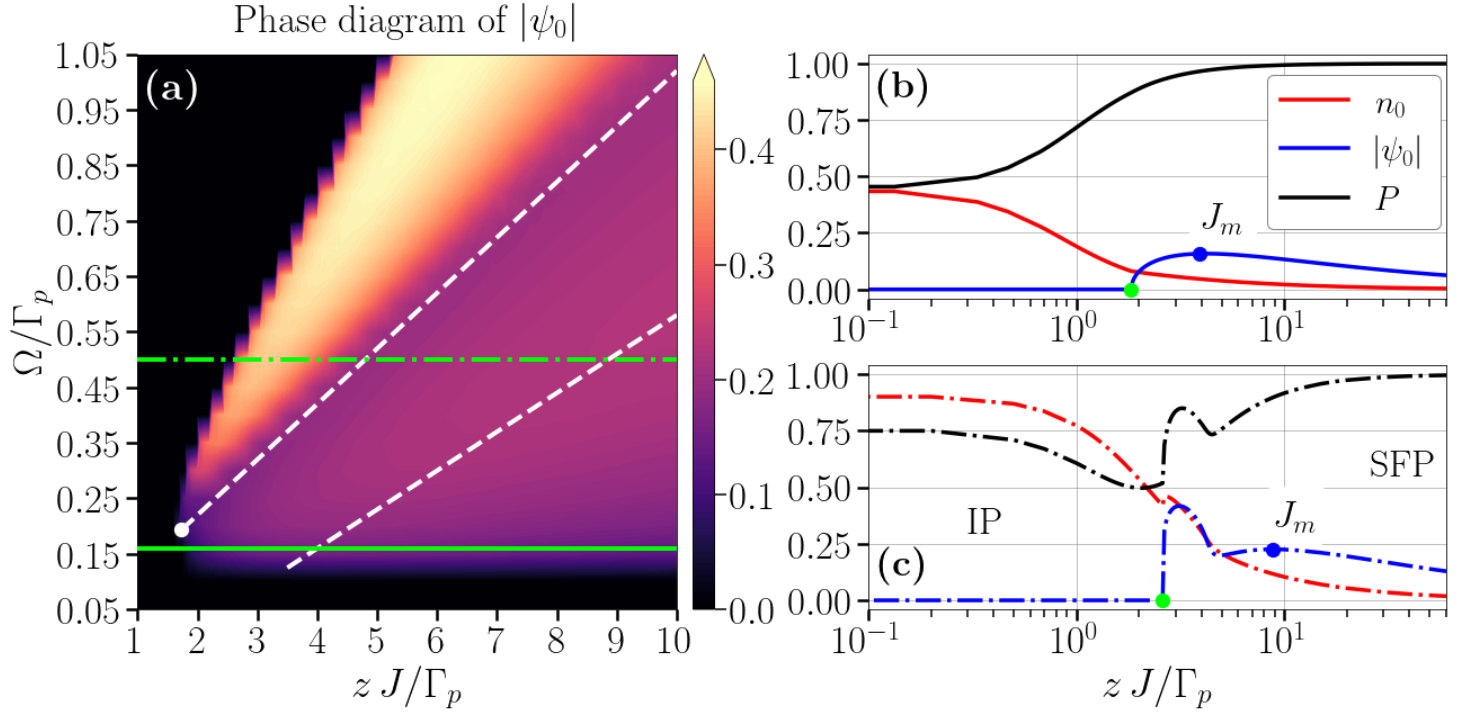}
    \caption{Top panel: sketch of the driven-dissipative system under consideration.
    (a) Mean-field phase diagram of the NESS for $\Gamma_l/\Gamma_p = 50 \, \gamma/\Gamma_p = 5 \cdot 10^{-2}$. The solid (dashed) green line corresponds to the horizontal cut at $\Omega/\Gamma_p = 1.6 \cdot 10^{-1}$ $\left( 5 \cdot 10^{-1} \right)$ shown in panel (b) [panel (c)]. The white point marks the tip of the SFP lobe, while the white dashed lines enclose the region of hole superfluidity.
    (b) Mean-field average density [red], order parameter [blue] and purity [black] across the Mott/superfluid transition at constant $\Omega/\Gamma_p = 1.6 \cdot 10^{-1}$ for the same parameters of panel (a). The green and blue dots highlight the critical point $J_c$ and the hopping scale $J_m$, respectively.
    (c) The equivalent of panel (b) for $\Omega/\Gamma_p = 5 \cdot 10^{-1}$.}
    \label{fig: phase_diagram}
\end{figure}

The evolution of the system is described by the Lindblad equation for the full density matrix $\hat{\rho}$,
\begin{equation}\label{eq: lindblad}
\begin{aligned}
    &\partial_t \hat{\rho} = -i \big[ \hat{H}_{\mathrm{BH}} + \hat{H}_{\mathrm{em}}, \hat{\rho} \big] \\
    &+ \frac{1}{2} \sum_{\mathbf{r}} \left( \Gamma_l \, \mathcal{D}{\big[ \hat{a}_{\mathbf{r}}; \hat{\rho} \big]} + \gamma \, \mathcal{D}{\big[ \hat{\sigma}^-_{\mathbf{r}}; \hat{\rho} \big]} + \Gamma_p \, \mathcal{D}{\big[ \hat{\sigma}^+_{\mathbf{r}}; \hat{\rho} \big]} \right) \, ,
\end{aligned}
\end{equation}
where $\mathcal{D}{\big[ \hat{O} \, ; \hat{\rho} \big]} = 2 \, \hat{O} \, \hat{\rho} \, \hat{O}^{\dagger} - \big\{ \hat{O}^{\dagger} \, \hat{O}, \hat{\rho} \big\}$.
For later convenience, we introduce the parameter $G \equiv \Omega^2/\left( \Gamma_p \, \Gamma_l \right)$, which defines the effective strength of the TLE-cavity coupling. A pictorial sketch of the model is provided in the top panel of Fig.~\ref{fig: phase_diagram}.

We study the NESS of the system within the Gutzwiller approximation~\cite{PhysRevLett.110.233601, PhysRevA.90.063821, PhysRevA.96.043809}. This consists in a site-factorized ansatz for the density matrix,
\begin{equation}\label{eq: GW_density_matrix}
    \hat{\rho} = \bigotimes_{\mathbf{r}} \sum_{n, m} \sum_{\sigma, \sigma'} c_{n, m, \sigma, \sigma'}{\left( \mathbf{r} \right)} \, | n, \sigma \rangle_{\mathbf{r}} \, \langle m, \sigma' |_{\mathbf{r}} \, ,
\end{equation}
where $| n, \sigma \rangle_{\mathbf{r}}$ is the local state of the cavity at $\mathbf{r}$ with $n$ cavity photons and TLE pseudospin $\sigma$. In this way, the Lindblad equation turns into a set of non-linear dynamical equations for the density matrix elements of the form $i \, \partial_t \, \vec{c}{\left( \mathbf{r} \right)} = \hat{L}{\left[ \vec{c}{\left( \mathbf{r} \right)} \right]} \cdot \vec{c}{\left( \mathbf{r} \right)}$, that we solve by numerical propagation until convergence to the homogeneous solution $\hat{\rho}_0 = \hat{\rho}{\left( t \to \infty \right)} = \vec{c}_0$~\cite{SM_2}. We stress that the ansatz~\eqref{eq: GW_density_matrix} entails a non-perturbative local description of the Rabi interaction~\eqref{eq: H_emitters}, which allows also to capture the physics of the so-called strong-coupling regime $G \gtrsim 1$.

In this paper, we restrict ourselves to the hard-core limit of our model $\left( U/J \to \infty \right)$, intended as an archetypal scenario of strong photon non-linearities. Moreover, we choose to set the TLE frequency to $\omega_{\mathrm{at}} = \omega_c - z \, J$ (with $z = 2 \, d$), in order that photons are explicitly pumped at the bottom of the cavity band and fragmentation effects due to finite-$\mathbf{k}$ condensation are avoided~\cite{PhysRevA.74.033612}. In this regime, the main properties of the NESS can be summarized as follows~\cite{SM_2}.

At fixed $\Omega$ and below a critical hopping $J_c$, the NESS is found to be in an insulating phase (IP) with vanishing order parameter $\psi_0 = \text{Tr}{\big( \hat{\rho}_0 \, \hat{a} \big)}$. In particular, for a large enough coupling $G \gg 1$ and $J \lesssim J_c$, the average photon density $n_0 = \text{Tr}{\big( \hat{\rho}_0 \, \hat{n} \big)}$ reaches a value close to 1, such that the cavity array hosts an essentially pure Mott insulating state~\cite{PhysRevA.96.023839, Ma2019}.
Increasing the cavity bandwidth $z \, J$, i.e. the kinetic energy, replenishment of lost photons occurs less efficiently, which leads to a substantial decrease in the density, alongside some entropy generation. At $J = J_c$, the NESS undergoes a second-order dynamical phase transition~\cite{PhysRevA.98.042118} to a superfluid phase (SFP), developing a finite order parameter displaying limit cycles $\psi_0 = \left| \psi_0 \right| e^{-i \, \omega_0 \, t}$ and scaling as $\left| \psi_0 \right| \sim \sqrt{J - J_c}$~\cite{PhysRevB.99.064511}. In the $\Omega-J$ projection of the phase diagram [Fig.~\ref{fig: phase_diagram}(a)], the SFP occupies a lobe-shaped region. We stress that here the formation of a coherent phase is not due to a competition between delocalization and local interactions as usual in strongly-correlated systems: instead, it is determined by the comparison between the emission bandwidth set by $\Gamma_p$ and the kinetic energy of holes propagating across the lattice.

In the SFP, $n_0$ is still an overall decreasing function of $J$, which acts similarly to a chemical potential for the system. Indeed, the oscillation or lasing frequency of the coherent field shows only a little deviation from its mean-field value at equilibrium, i.e. $\omega_0 \approx z \, J \left( 2 \, n_0 - 1 \right) + \omega_c$~\cite{SM_2}, meaning that the energy is lowered at large $J$ by depleting photons. On the other hand, the condensate density $\rho_c = \left| \psi_0 \right|^2$ is generally a non-monotonous function of $J$ and its behavior crucially depends on the value of $\Omega$. In the weak-coupling regime $G \sim 1$, located below the tip of the SFP lobe [Fig.~\ref{fig: phase_diagram}(b)], $\rho_c$ shows a maximum for $J = J_m$, after which it saturates $n_0$ to give an extremely pure and dilute condensate. This behavior can be understood as follows. For $J < J_m$, local losses $\Gamma_l$ become a non-negligible dissipation source and favour quantum coherence: since photon losses make the condensate density increase, the NESS can be classified as a \emph{hole superfluid}~\cite{krutitsky, BEC}. For $J > J_m$, the large bandwidth $z \, J$ overcomes the effect of all dissipative effects, so that cavity photons form a dilute \emph{particle superfluid}~\footnote{Formally, the notion of particle/hole superfluidity is related to $R = \text{sgn}{\left( \partial \rho_c/\partial n_0 \right)}$ at fixed $J$~\cite{krutitsky}. Nonetheless, we find that $R$ does not change if calculated at fixed $\Omega$ for the same values of $J$, see Sec.~\ref{SM-supsec: phase_diagram}(B) of the SM.}, whose purity increases with $J$.
Interestingly, in the strong-coupling regime $G \gtrsim 1$ [Fig.~\ref{fig: phase_diagram}(c)], a second maximum of $\rho_c$ develops for $J_c < J < J_m$, corresponding to a particle superfluid nearing the equilibrium hard-core state~\cite{SM_2}.

\textit{Collective excitations.}
Inspired by well-known linearization methods at equilibrium~\cite{krutitsky_navez, stringari2018}, we generalize the approach introduced in~\cite{PhysRevLett.110.233601} and consider small oscillations around the NESS configuration as
\begin{equation}\label{eq: GW_fluctuations}
    \vec{c}{\left( \mathbf{r}, t \right)} = \vec{c}_0 + \vec{u}_{\mathbf{k}} \, e^{i \left( \mathbf{k} \cdot \mathbf{r} - \omega_{\mathbf{k}} \, t \right)} + \vec{v}^*_{\mathbf{k}} \, e^{-i \left( \mathbf{k} \cdot \mathbf{r} - \omega^*_{\mathbf{k}} \, t \right)} \, ,
\end{equation}
as seen from the rotating frame of the coherent field~\cite{SM_2}.
Here, $\vec{u}_{\mathbf{k}} \left( \vec{v}_{\mathbf{k}} \right)$ weighs a particle (hole) excitation with energy $\omega_{\mathbf{k}} \left( -\omega^*_{\mathbf{k}} \right)$. Linearizing the Lindblad equation with respect to the fluctuations, one obtains the Bogoliubov-de Gennes equations
\begin{equation}\label{eq: GW_eigen}
    \omega_{\mathbf{k}}
    \begin{pmatrix}
    \vec{u}_{\mathbf{k}} \\
    \vec{v}_{\mathbf{k}}
    \end{pmatrix}
    = \hat{\mathcal{L}}_{\mathbf{k}} \begin{pmatrix}
    \vec{u}_{\mathbf{k}} \\
    \vec{v}_{\mathbf{k}}
    \end{pmatrix}
    \, ,
\end{equation}
where the superoperator $\hat{\mathcal{L}}_{\mathbf{k}} = \text{diag}{\big( \hat{A}_{\mathbf{k}}, -\hat{A}^*_{\mathbf{k}} \big)}$ is block-diagonal because of the relation $\vec{v}_{\mathbf{k}} = \left( \vec{u}_{\mathbf{k}} \right)^T$, due to the built-in Hermiticity of the density matrix~\cite{SM_2}. As a main result of this work, the eigenvalue equation~\eqref{eq: GW_eigen}
provides the energy spectra $\omega_{\alpha, \mathbf{k}}$ of the collective many-body excitations of the NESS
as well as
the strength of the response of the collective excitations to different perturbation channels. For instance, from the linearized expression of the photon density, $n{\left( \mathbf{r} \right)} = n_0 + \left( N_{\alpha, \mathbf{k}} \, e^{i \, \mathbf{k} \cdot \mathbf{r}} + \text{c.c.} \right)$, we can extract the spectral weight $N_{\alpha, \mathbf{k}} = \sum_{n, \sigma} n \, u_{n, n, \sigma, \sigma}$ of each mode in the density channel. Analogously, we can define the weight of particle (hole) excitations $U_{\alpha, \mathbf{k}} \left( V_{\alpha, \mathbf{k}} \right)$ through the fluctuations of the order parameter, $\psi{\left( \mathbf{r} \right)} = \psi_0 + U_{\alpha, \mathbf{k}} \, e^{i \, \mathbf{k} \cdot \mathbf{r}} + V^*_{\alpha, \mathbf{k}} \, e^{-i \, \mathbf{k} \cdot \mathbf{r}}$~\cite{SM_2}.



\begin{figure}
    \centering
    \includegraphics[width=1.0\linewidth]{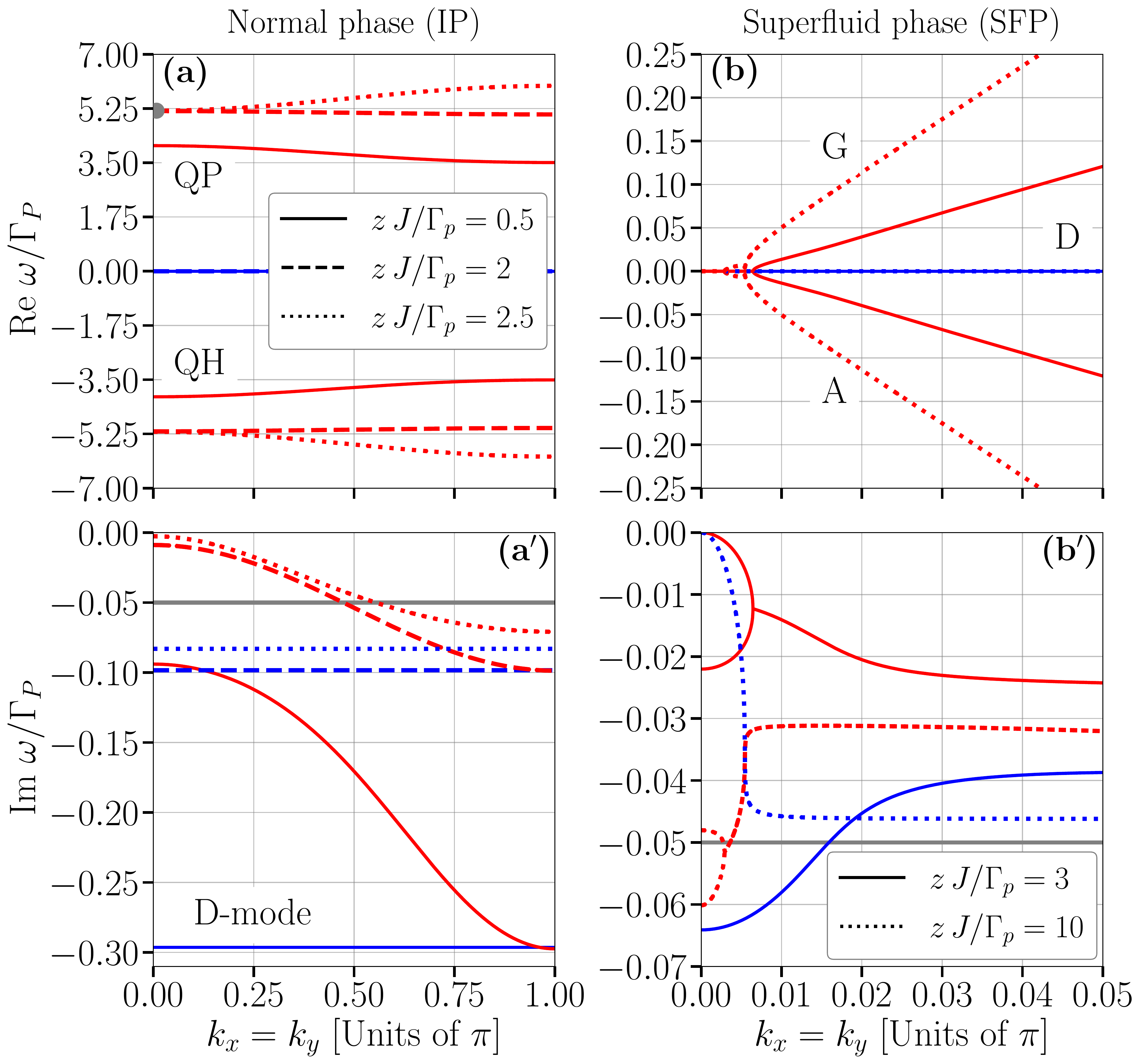}
    \caption{(a)-(a$'$) Excitation spectrum of the IP for $\Omega/\Gamma_p = 5 \cdot 10^{-1}$ [see Fig.~\ref{fig: phase_diagram}(c)] and increasing $z \, J/\Gamma_p$. The gray dot in panel (a) pinpoints the critical lasing frequency $\omega_*$. The acronym QP (QH) denotes the \quotes{quasiparticle} (\quotes{quasihole}) branch.
    (b)-(b$'$) Excitation spectrum of the SFP for $\Omega/\Gamma_p = 3 \cdot 10^{-1}$ and two values of $z \, J/\Gamma_p$ below/above the anti-adiabatic crossover at $J = J_m$. The letters G, A and D indicate the (standard) Goldstone, amplitude and D-mode branches.
    In panels (a$'$)-(b$'$), the gray solid line specifies the cavity loss rate $\Gamma_l/\Gamma_p$. All panels refer to a $d = 2$ array.}
    \label{fig: spectra}
\end{figure}

\textit{Insulating phase.}
The low-energy part~\footnote{The remaining excitations correspond to all the possible local modes of the system, with real energies proportional to suitable combinations of $\omega_c$ and $\omega_{\mathrm{at}}$, and damping rates of the order of $\Gamma_p$.} of the excitation spectrum in the IP phase [Fig.~\ref{fig: spectra}(a)-(a$'$)] consists of two dispersive branches $\omega_{\pm}{\left( \mathbf{k} \right)} = \pm \, \varepsilon_{\mathrm{ph}}{\left( \mathbf{k} \right)} - i \, \Gamma_{\mathrm{ph}}{\left( \mathbf{k} \right)}$ [red lines] and a purely dissipative local mode $\omega_{\mathrm{D}} = -i \, \Gamma_{\mathrm{D}}$ [blue lines], to which we refer as the \emph{D-mode}. The former bands correspond to distinct quasiparticle (quasihole) excitations respectively $\left( U_{-, \mathbf{k}} = V_{+, \mathbf{k}} = 0 \right)$, while the latter excites density fluctuations only. Deep in the IP phase [solid lines], the quasiparticle (QP) damping $\Gamma_{\mathrm{ph}}{\left( \mathbf{k} \right)}$ has a gapped, quadratic dispersion which extends up to the energy scale of the effective pumping rate $\Gamma_{\mathrm{em}} \simeq 4 \, \Omega^2/\Gamma_p = 4 \, \Gamma_l \, G$~\footnote{Strictly speaking, $\Gamma_{\mathrm{em}} = 4 \, \Omega^2/\Gamma_p$ is the effective pumping rate in the perturbative limit $\Omega \ll \Gamma_p$ (see e.g.~\cite{PhysRevA.96.023839, PhysRevA.96.033828}), however it provides a good rule of thumb for the order of magnitude of $\Gamma_{\mathrm{D}}$ also in the strong-coupling regime, see Sec.~\ref{SM-supsec: phase_diagram}(A) of the SM.}, indicating that non-local QP excitations at small $\left| \mathbf{k} \right|$ have a longer lifetime. As the hopping reaches the lasing threshold $J_c$ [dotted lines], the D-mode damping $\Gamma_{\mathrm{D}}$ approaches the bare cavity losses $\Gamma_l$ [gray solid line in Fig.~\ref{fig: spectra}(a$'$)], while the imaginary part of the longest-lived mode, corresponding to the so-called Liouvillian gap $\Gamma_{\mathrm{ph}}{\left( \mathbf{0} \right)}$ vanishes proportionally to $J_c - J$~\cite{PhysRevA.98.042118} as expected from quantum field theory~\cite{Sieberer_2016}: this substantiates the physical picture of long-lived QP's as precursors of the non-equilibrium transition to the SFP.

Tuning the hopping has a dramatic effect also on the real part of the QP excitation energy, which is well fitted by $\varepsilon_{\mathrm{ph}}{\left( \mathbf{k} \right)} \approx J{\left( \mathbf{k} \right)} \left( 1 - 2 \, n_0 \right) + \omega_c$ [where $J{\left( \mathbf{k} \right)}$ is the free-particle dispersion on the lattice] and is characterized by a density-dependent bandwidth. In detail, at small $J$ where $n_0 > 1/2$, $\varepsilon_{\mathrm{ph}}{\left( \mathbf{k} \right)}$ has an \emph{inverted} profile with minimal gap at $\mathbf{k} = \boldsymbol{\pi}$, while a more usual QP dispersion is found at larger $J$ when $n_0 < 1/2$. Eventually, $\varepsilon_{\mathrm{ph}}{\left( \mathbf{0} \right)}$ nears the lasing frequency $\omega_* = \omega^c_0$ [gray dot in Fig.~\ref{fig: spectra}(a)] at the transition point, which therefore can be regarded as an authentic finite-frequency criticality~\cite{RevModPhys.65.851, PhysRevB.99.064511}. This behavior finds an intuitive explanation in the aforementioned competition between hopping and dissipation. For $n_0 > 1/2$, photon pumping is efficient enough to prevent holes from moving around the hard-core lattice: thus, local particle-hole excitations are energetically favoured, despite their shorter lifetime. By contrast, in the opposite case the IP becomes hole-dominated and delocalized QP's are more likely to be excited. Importantly, we observe that there always exists a value of $J$ for which $n_0 = 1/2$, such that the QP band $\varepsilon_{\mathrm{ph}}{\left( \mathbf{k} \right)}$ is completely flat [dashed lines].


\textit{Superfluid phase.}
As the onset of the SFP corresponds to a spontaneous breaking of U(1) symmetry, the QP mode is replaced by a Goldstone branch $\omega_G{\left( \mathbf{k} \right)}$ whose dispersion vanishes in the long-wavelength limit~\cite{RevModPhys.65.851, stringari_pitaevskii}. Physically, this mode can be understood as a slow rotation of the condensate phase across the cavity array. Let us analyze the main features of different SFP regimes in more detail~\cite{SM_2}, starting from the region $J \lesssim J_m$.

For this case, a typical example of the excitation spectrum is given by the solid lines in Fig.~\ref{fig: spectra}(b)-(b$'$). Here, we recover in a novel strongly-correlated regime the usual behavior of out-of-equilibrium condensates with a diffusive $\text{Im} \, \omega_G{\left( \mathbf{k} \right)} \sim -\mathbf{k}^2$ and non-propagating Goldstone mode [red line with $\text{Re} \, \omega > 0$]~\cite{PhysRevLett.99.140402, Chiocchetta_2013, open_goldstones, PhysRevA.104.053516}.
Besides the Goldstone branch, we retrieve also its conjugate or \emph{ghost} mode $\omega_{\mathrm{A}}{\left( \mathbf{k} \right)}$ for negative energies~\cite{PhysRevLett.99.140402, PhysRevA.104.053516},
while the D-mode $\omega_{\mathrm{D}}{\left( \mathbf{k} \right)} \sim -i \, \Gamma_l$ acquires a non-trivial dispersion and retains a strong density character. The relationship between the latter mode and the Goldstone branch is pivotal to grasping the physics of the deep SFP, as we discuss in the following. For $J \lesssim J_m$, there is a clear scale separation between the imaginary parts of the Goldstone energy $\Gamma_G{\left( \mathbf{k} \right)}$ and the D-mode $\omega_{\mathrm{D}}{\left( \mathbf{k} \right)}$,
but such separation gets reduced for increasing $J$.


Indeed, the above situation changes dramatically at the boundary between particle and hole superfluidity $J = J_m$: here, the comparable time scales of pumping and loss processes cause the condensate to be dilute, so that the dissipative dynamics of the TLE's can no longer be adiabatically separated from that of the BH lattice.
This translates into a stable cross-hybridization of the D-mode $\omega_{\mathrm{D}}{\left( \mathbf{k} \right)}$ with the Goldstone branch $\Gamma_G{\left( \mathbf{k} \right)}$ at small momenta, which anyway leaves the real part of the energy spectrum unaltered [dashed lines in Fig.~\ref{fig: spectra}(b)-(b$'$)].

\textit{Dynamical response to a weak probe.}
Further light on the collective modes can be obtained from the linear response functions of the NESS, e.g. to an additional weak probe. A pioneering experiment of this family was reported in~\cite{Ma2019}, which suggests the experimental feasibility of our proposal. In particular, we focus on the retarded Green's function of the NESS to a one-particle coherent perturbation of the Hamiltonian \eqref{eq: BH_model}, whose Fourier-space form $G_R(\mathbf{k},\omega)$ directly provides
the \emph{transmission} $T{\left( \mathbf{k}, \omega \right)} = -i \, \Gamma_l \, G_{\mathrm{R}}{\left( \mathbf{k}, \omega \right)}$ and the \emph{reflection} $R{\left( \mathbf{k}, \omega \right)} = 1 + T{\left( \mathbf{k}, \omega \right)}$ amplitudes for a weak probe beam of wave vector $\mathbf{k}$ and frequency $\omega$~\cite{PhysRevA.74.033811, RevModPhys.85.299, chiocchetta2017}. For these experimentally accessible quantities, our theory provides the semi-analytical result $G_{\mathrm{R}}{\left( \mathbf{k}, \omega \right)} = \sum_{\alpha} Z_{\alpha, \mathbf{k}}/\left(\omega - \omega_{\alpha, \mathbf{k}} \right)$ with $Z_{\alpha, \mathbf{k}} \sim U_{\alpha, \mathbf{k}}$~\cite{SM_2}.

\begin{figure}
    \centering
    \includegraphics[width=1.0\linewidth]{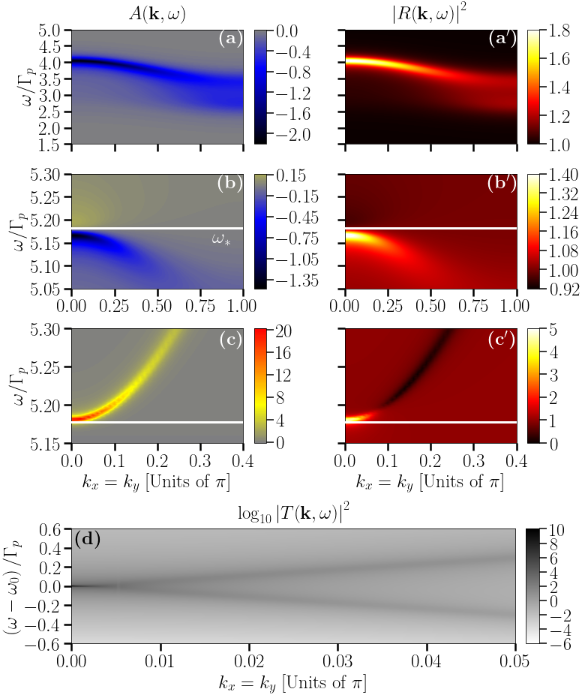}
    \caption{(a)-(c) DoS in the IP for the same parameters of Fig.~\ref{fig: spectra}(a)-(a$'$) and increasing $z \, J/\Gamma_p$ from top to bottom. (a$'$)-(c$'$) The related reflectivity spectrum. (d) Transmittivity in the SFP corresponding to the excitation spectra in Fig.~\ref{fig: spectra}(b)-(b$'$) for $z \, J/\Gamma_p = 10$ (anti-adiabatic regime). The white horizontal lines indicate the effective chemical potential $\omega_*$.}
    \label{fig: T_R}
\end{figure}

In the deep IP, we find that the transmittivity $\left| T{\left( \mathbf{k}, \omega \right)} \right|^2$~\cite{SM_2} displays a peak at the QP pole but remains well below unity.
The situation is strikingly different in the reflectivity channel [Fig.~\ref{fig: T_R}(a$'$)], which exhibits amplification as $\left| R{\left( \mathbf{k}, \omega \right)} \right|^2 > 1$. This result is tightly linked with the intrinsic out-of-equilibrium nature of the IP. While for the transmittivity we simply have $\left| T{\left( \mathbf{k}, \omega \right)} \right| \propto \left| G_{\mathrm{R}}{\left( \mathbf{k}, \omega \right)} \right|$, the reflectivity reads
\begin{equation}\label{eq: R}
    \left| R{\left( \mathbf{k}, \omega \right)} \right|^2 = \left[ 1 - \pi \, \Gamma^2_l \, A{\left( \mathbf{k}, \omega \right)} \right]^2 + \Gamma^4_l \left| \text{Re} \, G_{\mathrm{R}}{\left( \mathbf{k}, \omega \right)} \right|^2
\end{equation}
and is greater than 1 when the Density of States (DoS) $A{\left( \mathbf{k}, \omega \right)} \propto -\text{Im} \, G_{\mathrm{R}}{\left( \mathbf{k}, \omega \right)}$ is \emph{negative}, see Fig.~\ref{fig: T_R}(a). The DoS negativity, already observed in the presence of Markovian dissipation~\cite{PhysRevB.99.064511, scarlatella2019, PhysRevX.11.031018}, is a signature of the pump-induced population inversion taking place in the deep IP and is conventionally associated with energy gain~\cite{PhysRevA.74.063822, PhysRevLett.98.240601} but, at the same time, competes with the onset of macroscopic coherence. Thus, we can draw a Janus-faced portrait of the IP state: although behaving as an insulator from the viewpoint of its many-body excitations and lacking long-range coherence, its dynamical response is a precursor of  a lasing state, with a broadband amplification distributed along the QP dispersion.

Upon increasing $J$, the QP branch $\varepsilon_{\mathrm{ph}}{\left( \mathbf{k} \right)}$ is shifted to larger energies and gradually crosses the critical lasing frequency $\omega_*$. Interestingly, $\omega_*$
acts here as an effective chemical potential, as the DoS smoothly acquires a positive sign for $\omega > \omega_*$. This spectral redistribution strongly reflects on the IP response, especially when the QP band becomes flat [Fig.~\ref{fig: T_R}(b)-(b$'$)]. Whereas the transmittivity concentrates around the condensation point $\left( \mathbf{k = 0}, \omega_* \right)$, the reflectivity has a Fano-like shape around $\omega_*$: namely, $\left| R{\left( \mathbf{k}, \omega \right)} \right|^2$ is above (below) $1$ for $\omega < \omega_*$ $\left( > \omega_* \right)$, as a result of the sign flip of the DoS.

In proximity of the critical point [Fig.~\ref{fig: T_R}(c)-(c$'$)], the DoS is mostly positive and bounded by $\omega_*$ from below, while the divergence of $\text{Re} \, G_{\mathrm{R}}{\left( \mathbf{k}, \omega \right)}$ around $\omega_*$~\cite{PhysRevB.99.064511} marks the onset of condensation. This is visible as a sharp increase of the transmission and reflection response at low momenta even \emph{before} the transition. The only remaining trace of the compresence of high-energy QP states with a sizeable DoS and of the imminent onset of coherence, i.e. the first and second contributions to Eq.~\eqref{eq: R}, is a residual dark resonance of $\left| R{\left( \mathbf{k}, \omega \right)} \right|^2$, which eventually fades out at the critical point.

On the SFP side, the behavior of our strongly-interacting system differs from other out-of-equilibrium superfluids~\cite{PhysRevB.79.125311} in that the transmittivity and reflectivity are able to clearly resolve both the Goldstone and ghost branches [Fig.~\ref{fig: T_R}(d)]. In analogy with strongly-interacting BH superfluids at equilibrium~\cite{BEC}, this can be explained in terms of the emergent particle-hole symmetry of the long-wavelength SFP excitations, namely $\left| U_{\alpha, \mathbf{k}} \right| = \left| V_{\alpha, \mathbf{k}} \right|$ for each $\alpha$~\cite{SM_2}. However, the $\mathbf{k}$-space extension of the diffusive plateau of the Goldstone mode remains quite limited in the anti-adiabatic limit, and the mode is mostly visible as an enhancement of $\left| T{\left( \mathbf{k}, \omega_* \right)} \right|^2$ for $\mathbf{k \to 0}$. Since the density amplitude $N_{\mathrm{D}, \mathbf{k}}$ of the D-mode drastically changes across the anti-adiabatic crossover~\cite{SM_2}, we expect that a deeper insight into this regime could be obtained from dynamical observables probing the photon number statistics.


\textit{Conclusions.}
In this Letter, we have developed a Gutzwiller approach to the collective excitations of a driven-dissipative fluid of light in the regime of strong photon-photon interactions, focusing on the non-equilibrium Mott/superfluid transition of the system.
In particular, our results highlight experimentally accessible signatures of the surprising peculiarities of the non-equilibrium Mott state, shown to enable light amplification in a pump-and-probe configuration, and of the rich interplay between coherence and dissipation underlying the diffusive nature of the Goldstone mode. Thanks to its flexibility, our theory paves the way to a more general understanding of the exotic quantum phases that emerge in lattice systems driven out of equilibrium and can find experimental realization in the next generation of circuit-QED experiments.




\textit{Acknowledgements.} Stimulating discussions with M. Schirò, A. Biella, M. Seclì and M. Stefanini are warmly acknowledged. FC and MC acknowledge financial support from the Italian MIUR under the PRIN 2017 \quotes{CEnTraL} project (Prot. 20172H2SC4 005) and PRIN 2020 (Prot. 2020JLZ52N 002). IC acknowledges financial support from the European Union H2020-FETFLAG-2018-2020 project ``PhoQuS'' (n.820392), from the Provincia Autonoma di Trento, and from the Q@TN initiative.


\bibliographystyle{apsrev4-1}
\bibliography{refs}


\makeatletter\@input{work_file.tex}\makeatother


\end{document}


\title{Supplementary Material \texorpdfstring{\\}{} \quotes{Collective excitations of a strongly-correlated photon fluid \texorpdfstring{\\}{} stabilized by incoherent drive and dissipation}}

\author{Fabio Caleffi$^1$}
\email{bafioc11@gmail.com}
\author{Massimo Capone$^{1, 2}$}
\author{Iacopo Carusotto$^{3}$}
\affiliation{$^1$International School for Advanced Studies (SISSA), Via Bonomea 265, I-34136 Trieste, Italy}
\affiliation{$^2$CNR-IOM Democritos, Via Bonomea 265, I-34136 Trieste, Italy}
\affiliation{$^3$INO-CNR BEC Center and Dipartimento di Fisica, Universit\`a di Trento, Via Sommarive 14, I-38123 Povo, Italy}

\date{\today}
\maketitle


\onecolumngrid

\setcounter{equation}{0}
\setcounter{figure}{0}
\setcounter{table}{0}
\setcounter{page}{1}
\makeatletter
\renewcommand{\thesection}{S\arabic{section}}
\renewcommand{\theequation}{S\arabic{equation}}
\renewcommand{\thefigure}{S\arabic{figure}}

\section{From the mean-field Lindbladian \texorpdfstring{$\hat{L}{\left[ \vec{c} \right]}$}{} to steady-state expectation values}\label{supsec: mean_field_problem}
Adopting the vectorial representation of density matrices, also known as Choi-Jamiolkowski isomorphism~\cite{CHOI, JAMIOLKOWSKI}, the Lindblad equation transforms into a system of dynamical equations for the density matrix elements. The mean-field ansatz~\eqref{MAIN-eq: GW_density_matrix} allows to find an approximate solution to the non-equilibrium evolution at the price of introducing non-linearities. More precisely, the entries of the Gutzwiller-approximated Lindblad operator $\hat{L}{\left[ \vec{c}{\left( \mathbf{r} \right)} \right]}$ have the expression
\begin{equation}\label{eq: L0}
\begin{aligned}
    L^{n', m', \mu, \mu'}_{n, m, \sigma, \sigma'}{\left[ \vec{c}{\left( \mathbf{r} \right)} \right]} \equiv &-J \left[ \left( \sqrt{n' + 1} \, \delta_{n, n' + 1} \, \delta_{m, m'} \, \delta_{\sigma, \mu} \, \delta_{\sigma', \mu'} - \sqrt{m'} \, \delta_{n, n'} \, \delta_{m, m' - 1} \, \delta_{\sigma, \mu} \, \delta_{\sigma, \mu} \right) \sum_{\mathbf{s}{\left( \mathbf{r} \right)}} \psi{\left( \mathbf{s} \right)} \right. \\
    &\left. + \left( \sqrt{n'} \, \delta_{n, n' - 1} \, \delta_{m, m'} \, \delta_{\sigma, \mu} \, \delta_{\sigma', \mu'} - \sqrt{m' + 1} \, \delta_{n, n'} \, \delta_{m, m' + 1} \, \delta_{\sigma, \mu} \, \delta_{\sigma', \mu'} \right) \sum_{\mathbf{s}{\left( \mathbf{r} \right)}} \psi^*{\left( \mathbf{s} \right)} \right] \\
    &+ \left\{ \omega_c \left( n' - m' \right) + U \left[ n' \left( n' - 1 \right) - m' \left( m' - 1 \right) \right] + \omega_{\mathrm{at}} \, \frac{\mu - \mu'}{2} \right\} \delta_{n, n'} \, \delta_{m, m'} \, \delta_{\sigma, \mu} \, \delta_{\sigma', \mu'} \\
    &+ \Omega \left( \sqrt{n' + 1} \, \delta_{n, n' + 1} \, \delta_{m, m'} \, \delta_{\sigma, -1} \, \delta_{\sigma, -\mu} \, \delta_{\sigma', \mu'} - \sqrt{m'} \, \delta_{n, n'} \, \delta_{m, m' - 1} \, \delta_{\sigma', 1} \, \delta_{\sigma, \mu} \, \delta_{\sigma', -\mu'} \right. \\
    &\left. + \sqrt{n'} \, \delta_{n, n' - 1} \, \delta_{m, m'} \, \delta_{\sigma, 1} \, \delta_{\sigma, -\mu} \, \delta_{\sigma', \mu'} - \sqrt{m' + 1} \, \delta_{n, n'} \, \delta_{m, m' + 1} \, \delta_{\sigma', -1} \, \delta_{\sigma, \mu} \, \delta_{\sigma', -\mu'} \right) \\
    &+ i \, \Gamma_l \left[ \sqrt{n' \, m'} \, \delta_{n, n' - 1} \, \delta_{m, m' - 1} \, \delta_{\sigma, \mu} \, \delta_{\sigma', \mu'} - \frac{1}{2} \left( n' + m' \right) \delta_{n, n'} \, \delta_{m, m'} \, \delta_{\sigma, \mu} \, \delta_{\sigma', \mu'} \right] \\
    &+ i \, \gamma \left( \delta_{\sigma, -1} \, \delta_{\sigma', -1} \, \delta_{n, n'} \, \delta_{m, m'} \, \delta_{\sigma, -\mu} \, \delta_{\sigma', -\mu'} - \frac{2 + \mu + \mu'}{4} \, \delta_{n, n'} \, \delta_{m, m'} \, \delta_{\sigma, \mu} \, \delta_{\sigma', \mu'} \right) \\
    &+ i \, \Gamma_p \left( \delta_{\sigma, 1} \, \delta_{\sigma', 1} \, \delta_{n, n'} \, \delta_{m, m'} \, \delta_{\sigma, -\mu} \, \delta_{\sigma', -\mu'} - \frac{2 - \mu - \mu'}{4} \, \delta_{n, n'} \, \delta_{m, m'} \, \delta_{\sigma, \mu} \, \delta_{\sigma', \mu'} \right) \, ,
\end{aligned}
\end{equation}
where the notation $\mathbf{s}{\left( \mathbf{r} \right)}$ labels the nearest-neighbouring sites of $\mathbf{r}$, the indices $n, n', m, m'$ and $\sigma, \sigma = \pm 1$ run over the occupation numbers of the cavity modes and the TLE pseudospin states respectively, and
\begin{equation}\label{eq: psi_formula}
    \psi{\left( \mathbf{r} \right)} \equiv \text{Tr}{\left( \hat{\rho} \, \hat{a}_{\mathbf{r}} \right)} = \sum_n \sum_{\sigma} \sqrt{n + 1} \, c_{n + 1, n, \sigma, \sigma}{\left( \mathbf{r} \right)}
\end{equation}
is the superfluid order parameter, namely the condensate amplitude. The mean-field stationary state of the system $\vec{c}_0$ is determined either by propagating in time the Gutzwiller-Lindblad equation $i \, \partial_t \, \vec{c}{\left( \mathbf{r} \right)} = \hat{L}{\left[ \vec{c}{\left( \mathbf{r} \right)} \right]} \cdot \vec{c}{\left( \mathbf{r} \right)}$ through a fourth-order Runge-Kutta algorithm until it converges to the NESS or by directly solving the linear problem for the NESS.

Besides the order parameter value, other local expectations values are straightforwardly evaluated through the following formulas,
\begin{equation}\label{eq: n_formula}
    n{\left( \mathbf{r} \right)} \equiv \text{Tr}{\left( \hat{\rho} \, \hat{n}_{\mathbf{r}} \right)} \equiv \sum_n \sum_{\sigma} n \, c_{n, n, \sigma, \sigma}{\left( \mathbf{r} \right)} \, ,
\end{equation}
\begin{equation}\label{eq: delta_n_formula}
    \Delta n^2{\left( \mathbf{r} \right)} \equiv \text{Tr}{\left\{ \hat{\rho} \left[ \hat{n}_{\mathbf{r}} - n{\left( \mathbf{r} \right)} \right]^2 \right\}} = \sum_n \sum_{\sigma} n^2 \, c_{n, n, \sigma, \sigma}{\left( \mathbf{r} \right)} - n{\left( \mathbf{r} \right)}^2 \, ,
\end{equation}
\begin{equation}\label{eq: s_z_formula}
    S_z{\left( \mathbf{r} \right)} \equiv \frac{1}{2} \text{Tr}{\left( \hat{\rho} \, \hat{\sigma}^z_{\mathbf{r}} \right)} = \frac{1}{2} \sum_n \sum_{\sigma} \sigma \, c_{n, n, \sigma, \sigma}{\left( \mathbf{r} \right)} \, ,
\end{equation}
\begin{equation}\label{eq: s_m_formula}
    S_-{\left( \mathbf{r} \right)} \equiv \frac{1}{2} \text{Tr}{\left( \hat{\rho} \, \hat{\sigma}^-_{\mathbf{r}} \right)} = \frac{1}{2} \sum_n c_{n, n, 1, -1}{\left( \mathbf{r} \right)} \, ,
\end{equation}
providing respectively the average photon density (namely, the average occupation number of the cavity mode) and its variance, and the population difference and polarization of the TLE coupled to the cavity (not examined in this work) at site $\mathbf{r}$. Moreover, we can easily estimate the purity
\begin{equation}\label{eq: P_formula}
    P \equiv \text{Tr}{\left( \hat{\rho}^2 \right)} = \sum_{n, m} \sum_{\sigma, \sigma'} \left| c_{n, m, \sigma, \sigma'} \right|^2
\end{equation}
and the von-Neumann entropy
\begin{equation}\label{eq: S_formula}
    S \equiv -\text{Tr}{\left[ \hat{\rho} \, \ln{\left( \hat{\rho} \right)} \right]} = -\sum_i \lambda_i \, \ln{\left( \lambda_i \right)}
\end{equation}
of the NESS, where we have made the hypothesis of a uniform steady state under our local approximation of $\hat{\rho}$. In particular, we notice that the purity $P$ is identical to the Frobenius norm of the density matrix. In Eq.~\eqref{eq: S_formula}, the summation runs over the eigenvalues $\lambda_i$ of the density matrix $\hat{\rho}$.


\section{Details on the phase diagram of the hard-core stationary state}\label{supsec: phase_diagram}
\subsection{Approximate estimation of the Mott/superfluid critical boundary}\label{supsubsec: critical_boundary}
Quantitative insights on the energy scales involved in the development of coherence inside the IP and in the breaking of U(1) symmetry in the SFP can be obtained by an exact derivation of the NESS density matrix $\vec{c}_0$. Formally, this calculation amounts to simply determine the unique eigenvector of $\hat{L}{\left[ \vec{c}{\left( \mathbf{r} \right)} \right]}$ with vanishing eigenvalue, which corresponds to the mean-field approximation of the IP state. As a final result, in the hard-core limit $U/J \to \infty$ (where the bosonic occupation number is restricted to $0, 1$) we find that the density matrix consists of five independent coefficients only, having expressions
\begin{subequations}
    \begin{equation}
        \left( c_0 \right)_{0, 1, 1, -1} = \frac{z \, J \, \Gamma_l}{\Gamma_p \, \Omega} + i \frac{\left( \Gamma_p + \Gamma_l + \gamma \right) \Gamma_l}{2 \, \Gamma_p \, \Omega} \, ,
    \end{equation}
    \begin{equation}
        \left( c_0 \right)_{0, 0, -1, -1} = \frac{\gamma}{\Gamma_p} c_{0, 0, 1, 1} + \frac{\left( \Gamma_l + \gamma \right) \Gamma_l}{\Gamma^2_p} \, ,
    \end{equation}
    \begin{equation}
        \left( c_0 \right)_{0, 0, 1, 1} = \frac{\gamma}{\Gamma_p} + \frac{\Gamma_p \, \Gamma_l}{4 \, \Omega^2} \left[ \left( \frac{2 \, \Omega}{\Gamma_p} \right)^2 + \left( 1 + \frac{\Gamma_l + \gamma}{\Gamma_p} \right)^2 + \left( \frac{2 \, z \, J}{\Gamma_p} \right)^2 \right] \, ,
    \end{equation}
    \begin{equation}
        \left( c_0 \right)_{1, 1, -1, -1} = \frac{\Gamma_l + \gamma}{\Gamma_p} \, ,
    \end{equation}
    \begin{equation}
        \left( c_0 \right)_{1, 1, 1, 1} = 1 \, ,
    \end{equation}
\end{subequations}
to be normalized by the trace $\text{Tr}{\left( \hat{\rho} \right)} = \sum_n \sum_{\sigma} \left( c_0 \right)_{n, n, \sigma, \sigma}$. Now, assuming to work in the limit $\gamma \ll \Gamma_p$, so that the TLE's are most of the time in their excited state $\sigma = 1$, and under the realistic condition $\Gamma_l \ll \Gamma_p$, the order of magnitude of the critical value of the photon bandwidth $z \, J_c$ can be straightforwardly obtained by imposing
\begin{equation}\label{J_c_condition}
    \frac{\left( c_0 \right)^{J = J_c}_{0, 0, 1, 1}}{\left( c_0 \right)^{J = J_c}_{1, 1, 1, 1}} \simeq 1 \, ,
\end{equation}
namely that the probability of a cavity to be either occupied or empty is about the same. From this condition, we get
\begin{equation}\label{critical_J}
    \frac{z \, J_c}{\Gamma_p} \simeq \frac{1}{2} \sqrt{\frac{\Gamma_{\mathrm{em}}}{\Gamma_l} - 1} \, ,
\end{equation}
where $\Gamma_{\mathrm{em}} = 4 \, \Omega^2/\Gamma_p$ is the effective photon emission rate into the cavities and which is in good agreement with the location of the Mott/superfluid boundary for a moderately large $\Omega$. Importantly, we note that the critical point exists as long as $\Gamma_{\mathrm{em}}/\Gamma_l \ge 1$, namely when the light emission rate exceeds losses, which sets a \textit{lasing condition} for the stabilization of the superfluid state. Therefore, $\Gamma_{\mathrm{em}}$ can be regarded as a reliable estimation of the effective pumping rate as stated in the main text. Instructive information can be also extracted from the off-diagonal element $\left( c_0 \right)_{0, 1, 1, -1}$, which quantifies the degree of quantum coherence in the IP. Comparing its (unnormalized) modulus at $J = 0$ with the value at $J = J_c$, we find
\begin{equation}
    \left| \left( c_0 \right)^{J = 0}_{0, 1, 1, -1} \right| = \frac{\left( \Gamma_p + \Gamma_l + \gamma \right) \Gamma_l}{2 \, \Gamma_p \, \Omega} < \left| \left( c_0 \right)^{J = J_c}_{0, 1, 1, -1} \right| \simeq \sqrt{\frac{\Gamma_l}{\Gamma_p}}
\end{equation}
where the inequality holds compatibly with the usual hypothesis $\Gamma_p \gg \Gamma_l, \gamma$ and, not surprisingly, when the lasing condition $\Gamma_{\mathrm{em}}/\Gamma_l \ge 1$ is fulfilled. Therefore, coherence in the IP state develops upon increasing the hopping and strongly relates to the photon leakage rate $\Gamma_l$ around the critical point, as we also observed by commenting the excitation spectrum of the IP in the main text. This result points out once more the key role played by different dissipation mechanisms in building coherence in the system.

\subsection{Additional features of the phase diagram}\label{supsubsec: additional_MF_features}

\begin{figure}[!t]
    \centering
    \includegraphics[width=0.8\linewidth]{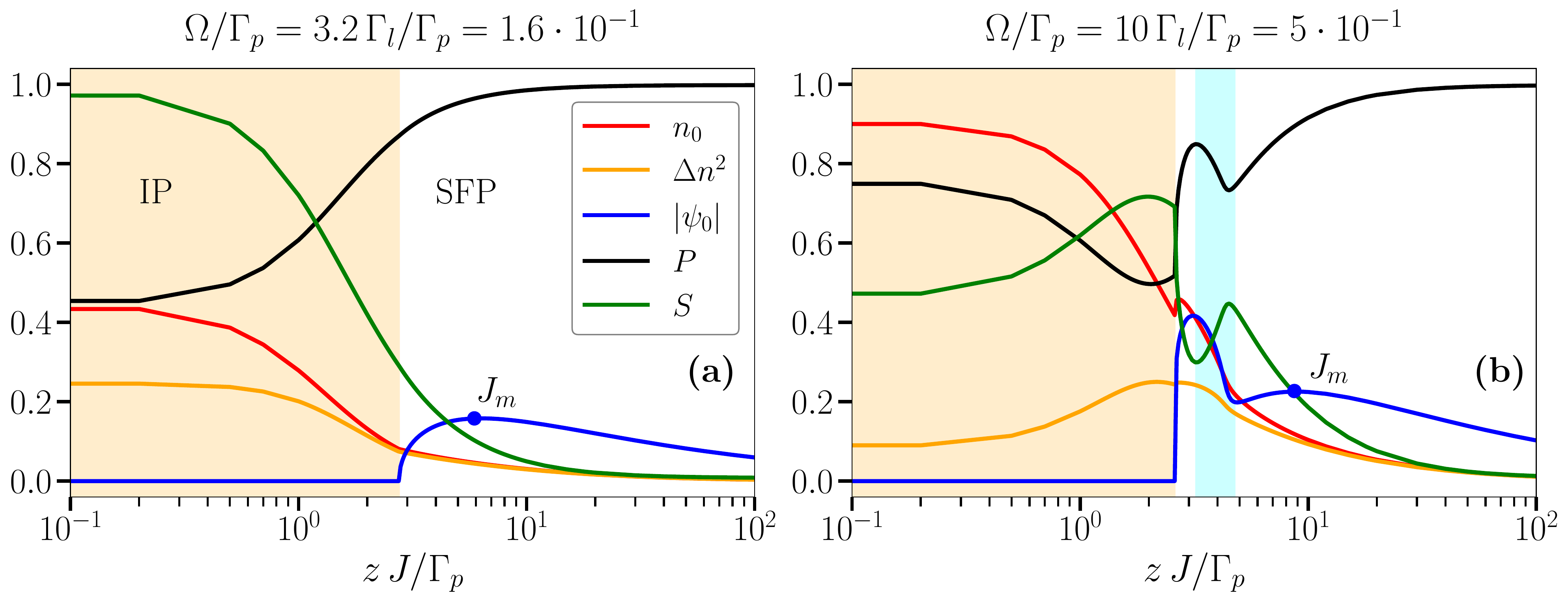}
    \caption{Panel (a): mean-field behavior of the NESS observables across the Mott/superfluid transition at constant $\Omega/\Gamma_p = \Gamma_l/\Gamma_p = 1.6 \cdot 10^{-1}$ and $\gamma/\Gamma_p = 10^{-3}$ in the weak-coupling regime, see also the solid lines in Fig.~\ref{MAIN-fig: phase_diagram}(b) of the main text.
    Panel (b): the equivalent of panel (a) at a larger value of $\Omega/\Gamma_p = 5 \cdot 10^{-1}$ in the strong-coupling regime, see also the dot-dashed lines in Fig.~\ref{MAIN-fig: phase_diagram}(b) of the main text.
    Color code: average photon density [red]; density variance [orange]; cavity order parameter [blue]; purity [black]; entropy [green]. The white (orange-shaded) area indicates the SFP (IP) region, while the cyan-shaded stripe in panel (b) highlights the values of $z \, J$ for which a dynamical instability due to the Goldstone mode appears, see Subsection~\ref{supsec: sf_excitations}(B).}
    \label{fig: figure_1_SM}
\end{figure}

In this Subsection, we give a more detailed account of the mean-field properties of the NESS that we touch upon in the main text discussion.

In Fig.~\ref{fig: figure_1_SM}, we provide a complete view of the mean-field predictions for local observables across the phase diagram in the weak-coupling [panel (a)] and strong-coupling [panel (b)] regimes, corresponding to TLE-cavity couplings $G \sim 1$ and $G \gtrsim 1$ respectively, where $G = \Omega^2/\left( \Gamma_p \, \Gamma_l \right)$. In the former case, the entropy $S$ (purity $P$) of the NESS is maximal (minimal) in the deep IP: here, the effective pumping is not sufficiently large compared to photon losses to stabilize a Mott-like state, but leads to a mixed state with a large density variance. The onset of coherence by the introduction of hopping processes purifies the NESS, as the average density rapidly approaches conditions of diluteness. Quite different behaviors and non-monotonicities appear instead in the strong-coupling regime, as already pointed out in the main text. In this case, the entropy (purity) is an increasing (decreasing) function of $J$ in the IP, except for a neighbourhood of the Mott/superfluid critical point. In the SFP, the behaviors of $S$ and $P$ change abruptly and follow again symmetrically opposite trends, with the former reaching a minimum where the order parameter develops its absolute maximum. We finally notice that, in the strong-coupling regime, the density variance is maximal in the vicinity of the Mott/superfluid transition; notably, we also find that the maximal value of the condensate density is remarkably close to its equilibrium value $\rho_c = \Delta n^2/n_0$, a quantitative match that further supports our description of the strongly-coupled SFP at optimal $J$ as the closest realization of the hard-core bosonic state at equilibrium found in our system.

\begin{figure}[!t]
    \centering
    \includegraphics[width=1.0\linewidth]{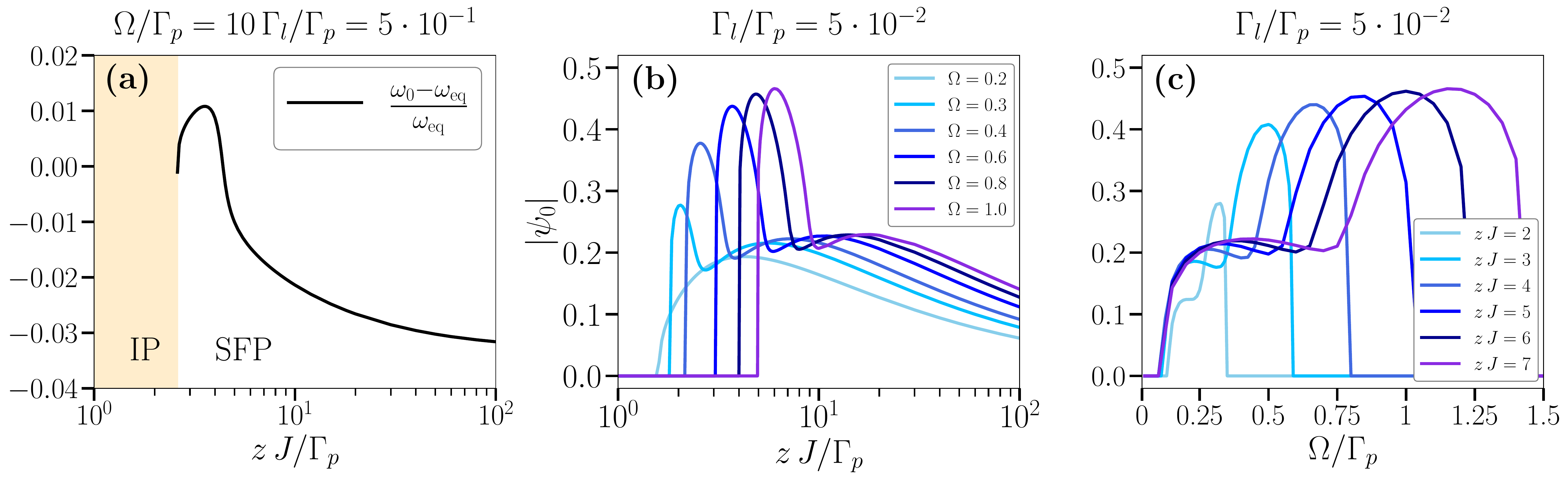}
    \caption{Panel (a): $J$-dependence of the relative deviation of the lasing frequency of the superfluid order parameter from its equilibrium value $\omega_{\mathrm{eq}} = z \, J \left( 2 \, n_0 - 1 \right) + \omega_c$ for $\Omega/\Gamma_p = 5 \cdot 10^{-1}$, $\Gamma_l/\Gamma_p = 5 \cdot 10^{-2}$ and $\gamma/\Gamma_p = 10^{-3}$. The white (orange-shaded) area indicates the SFP (IP) region.
    Panel (b): $J$-dependence of the order parameter of the SFP for different constant values of $\Omega$, ranging from the weak-coupling regime (light-blue line) to the strong-coupling limit (purple line), corresponding to distinct horizontal cuts of the phase diagram shown in Fig.~\ref{MAIN-fig: spectra}(a) of the main text.
    Panel (c): $\Omega$-dependence of the order parameter of the SFP for increasing constant values of $J$, corresponding to distinct vertical cuts of the same phase diagram.}
    \label{fig: figure_2_SM}
\end{figure}

Turning our attention specifically to the SFP, we first comment on the oscillation or lasing frequency $\omega_0$ of the order parameter, the typical behavior as a function of the hopping $J$ is shown in the leftmost panel of Fig.~\ref{fig: figure_2_SM}. Here, we show the deviation of $\omega_0$ from the effective chemical potential of the hard-core system at equilibrium $\omega_{\mathrm{eq}} = z \, J \left( 2 \, n_0 - 1 \right) + \omega_c$, see also Subsection~\ref{supsec: equilibrium_hard_core}(B). $\omega_0$ is exactly equal to its equilibrium counterpart at the critical point $J_c$ and displays a highly non-monotonic behavior across the various regimes of the SFP. Eventually, the relative deviation $\left( \omega_0 - \omega_{\mathrm{eq}} \right)/\omega_{\mathrm{eq}}$ converges to a finite value in the dilute limit of the system.

A more intriguing property of the SFP is represented by its particle/hole character, which we addressed by comparing the different behaviours of the condensate density $\rho_c = \left| \psi_0 \right|^2$ and the cavity filling $n_0$ either at fixed $J$ or at fixed $\Omega$. In Fig.~\ref{fig: figure_2_SM}(b) and (c), we illustrate the behavior of the order parameter modulus $\left| \psi_0 \right|$ as a function of $J$ and $\Omega$ respectively, showing in particular how the strong-coupling behavior discussed in the main text gradually develops upon increasing $\Omega$ in the former panel. In particular, we verify that the SFP regions where $\left| \psi_0 \right|$ is an increasing function of $J$ at constant $\Omega$ coincide exactly with the regions where $\left| \psi_0 \right|$ is a decreasing function of $\Omega$ at constant $J$. In other words, $J$ and $\Omega$ turn out to have opposite physical roles from the point of view of the superfluid order parameter. This justifies the diagonal orientation of the hole-superfluidity area highlighted in the mean-field phase diagram reported in Fig.~\ref{MAIN-fig: phase_diagram}(a) of the main text. Importantly, the same observation applies also to $n_0$, which is always an overall decreasing (increasing) function of $J$ ($\Omega$). It immediately follows that the particle/hole superfluid character $R = \text{sgn}{\left( \partial \rho_c/\partial n_0 \right)}$ does not depend on which quantity is held fixed while evaluating the derivative.

\section{Mean-field theory and collective excitations of the hard-core Bose gas}\label{supsec: equilibrium_hard_core}
\subsection{Holstein-Primakoff mapping}\label{supsubsec: hp_mapping}
In the hard-core limit $U/J \to \infty$, the BH model~\eqref{MAIN-eq: BH_model} can be recast into the Hamiltonian of a XXZ model through the Holstein-Primakoff mapping of hard-core bosons to spin operators $\hat{\tau}^i_{\mathbf{r}}$. In particular, under the identifications $\hat{a}_{\mathbf{r}} \to \hat{\tau}^-_{\mathbf{r}}$ and $\hat{n}_{\mathbf{r}} \to \left( \hat{\tau}^z_{\mathbf{r}} + 1 \right)/2$, we obtain
\begin{equation}\label{eq: XXZ_model}
    \hat{H}_{\mathrm{BH}} = -J \sum_{\langle \mathbf{r}, \mathbf{s} \rangle} \hat{\tau}^+_{\mathbf{r}} \, \hat{\tau}^-_{\mathbf{s}} + \frac{\bar{U}}{4} \sum_{\langle \mathbf{r}, \mathbf{s} \rangle} \hat{\tau}^z_{\mathbf{r}} \, \hat{\tau}^z_{\mathbf{s}} + \frac{2 \, \omega_c + z \, \bar{U}}{4} \sum_{\mathbf{r}} \left( \hat{\tau}^z_{\mathbf{r}} + 1 \right) \, ,
\end{equation}
where we have also included a nearest-neighbour interaction term with energy scale $\bar{U}$ for the sake of generality. More in detail, this is exactly the non-local repulsive coupling that emerges from second-order perturbation theory on top of the Mott state of the model when one considers a very large but finite value of the Hubbard interaction energy $U$: in this case, $\bar{U} \propto t^2/U$. The equations of motion of the spin operators can be readily found via the Heisenberg equation, yielding
\begin{equation}\label{eq: tau_z}
    i \, \dot{\hat{\tau}}^z_{\mathbf{r}} = -2 \, J \, \sum_{\mathbf{s}{\left( \mathbf{r} \right)}} \left( \hat{\tau}^+_{\mathbf{r}} \, \hat{\tau}^-_{\mathbf{s}} - \text{h.c.} \right) \, ,
\end{equation}
\begin{equation}\label{eq: tau_-}
    i \, \dot{\hat{\tau}}^-_{\mathbf{r}} = J \, \hat{\tau}^z_{\mathbf{r}} \sum_{\mathbf{s}{\left( \mathbf{r} \right)}} \hat{\tau}^-_{\mathbf{s}} + \frac{\bar{U}}{2} \, \hat{\tau}^-_{\mathbf{r}} \sum_{\mathbf{s}{\left( \mathbf{r} \right)}} \hat{\tau}^z_{\mathbf{s}} + \frac{2 \, \omega_c + z \, \bar{U}}{2} \, \hat{\tau}^-_{\mathbf{r}} \, .
\end{equation}

\subsection{Mean-field theory at equilibrium}\label{supsubsec: hard_core_mean_field_theory}
Within the mean-field approximation, we consider the average values of the operators appearing in Eqs.~\eqref{eq: tau_z}-\eqref{eq: tau_-} and decouple those fields that act on different sites. Renaming $\langle \hat{\tau}^z_{\mathbf{r}} \rangle \to 2 \, n{\left( \mathbf{r} \right)} - 1$ and $\langle \hat{\tau}^-_{\mathbf{r}} \rangle \to \psi{\left( \mathbf{r} \right)}$ (which stands for the photonic order parameter), we have
\begin{equation}\label{eq: mf_hc_n_eq}
    i \, \dot{n}{\left( \mathbf{r} \right)} = -J \, \sum_{\mathbf{s}{\left( \mathbf{r} \right)}} \left[ \psi^*{\left( \mathbf{r} \right)} \, \psi{\left( \mathbf{s} \right)} - \text{c.c.} \right] \, ,
\end{equation}
\begin{equation}\label{eq: mf_hc_psi_eq}
    i \, \dot{\psi}{\left( \mathbf{r} \right)} = 2 \, J \, n{\left( \mathbf{r} \right)} \sum_{\mathbf{s}{\left( \mathbf{r} \right)}} \psi{\left( \mathbf{s} \right)} + \bar{U} \, \psi{\left( \mathbf{r} \right)} \sum_{\mathbf{s}{\left( \mathbf{r} \right)}} n{\left( \mathbf{s} \right)} - J \sum_{\mathbf{s}{\left( \mathbf{r} \right)}} \psi{\left( \mathbf{s} \right)} + \omega_c \, \psi{\left( \mathbf{r} \right)} \, ,
\end{equation}
which can be regarded as the Gross-Pitaevskii equations (GPE) for the hard-core regime. Let us now assume that the stationary state of the system is provided by a uniform ansatz for the average local density $n{\left( \mathbf{r} \right)} = n_0$ and for the order parameter $\psi{\left( \mathbf{r} \right)} = \psi_0 \, e^{-i \, \omega \, t}$. Therefore, we are left with only one equation,
\begin{equation}\label{eq: mf_hc_psi_ansatz}
    \omega \, \psi_0 = z \, J \left( 2 \, n_0 - 1 \right) \psi_0 + z \, \bar{U} \, n_0 \, \psi_0 + \omega_c \, \psi_0 \, ,
\end{equation}
providing the oscillation frequency of the order parameter,
\begin{equation}\label{eq: mf_hc_psi_omega}
    \omega_{\mathrm{eq}} = z \, J \left( 2 \, n_0 - 1 \right) + z \, \bar{U} \, n_0 + \omega_c \, ,
\end{equation}
which can be also regarded as the effective chemical potential of a quantum degenerate gas~\cite{stringari_pitaevskii}. In particular, we point out that, in the case of the hard-core lattice, the mean-field shift of the energy due to interactions is solely due to the hopping and reads $\Delta \mu = z \left( 2 \, J + \bar{U} \right) n_0$. Notice also that a special property of the hard-core regime is the coupling between the density and order parameter fields. A noteworthy consequence of this fact is the exact relation $\left| \psi{\left( \mathbf{r} \right)} \right|^2 = n{\left( \mathbf{r} \right)} \left[ 1 - n{\left( \mathbf{r} \right)} \right]$ valid within mean-field theory at equilibrium, meaning that the order parameter modulus is the geometric average of the particle and hole densities.

Taking advantage of the identity $\left| \psi_0 \right|^2 = n_0 \left( 1 - n_0 \right)$, we can calculate the optimal value of $n_0$ through the minimization of the mean-field energy
\begin{equation}\label{eq: mf_hc_energy}
    E_{\mathrm{BH}}/V = -z \, J \, \left| \psi_0 \right|^2 + \frac{z \, \bar{U} \, n^2_0}{2} + \omega_c \, n_0 \, ,
\end{equation}
where $V$ is the lattice volume. In particular, we obtain
\begin{equation}\label{eq: mf_hc_n}
    n_0 = \frac{z \, J - \omega_c}{z \left( 2 \, J + \bar{U} \right)} \, ,
\end{equation}
so that
\begin{equation}\label{eq: mf_hc_psi}
    \left| \psi_0 \right|^2 = \frac{\left( z \, J - \omega_c \right) \left[ z \left( J + \bar{U} \right) + \omega_c \right]}{z^2 \left( 2 \, J + \bar{U} \right)^2} \, .
\end{equation}
With reference to the strong-coupling regime of our driven-dissipative model, we find that the more pure is the hard-core state approached by the SFP for $J_c < J < J_m$ (where $J_m$ marks the crossover to the dilute regime of the SFP), the closer the average density is to the half-filling value $n_0 = 1/2$, see e.g. \autoref{fig: figure_1_SM}(b).

\subsection{Collective excitations of the superfluid state at equilibrium}\label{supsubsec: hard_core_modes}
The elementary excitations on top of the mean-field hard-core state determined before can be accessed by considering small-amplitude fluctuations around the condensate order parameter and the average density as follows,
\begin{equation}\label{eq: fluctuations_n_eq}
    n{\left( \mathbf{r} \right)} = n_0 \left[ 1 + w_{\mathbf{k}} \, e^{i \left( \mathbf{k} \cdot \mathbf{r} - \omega_{\mathbf{k}} \, t \right)} + w^*_{\mathbf{k}} \, e^{-i \left( \mathbf{k} \cdot \mathbf{r} - \omega_{\mathbf{k}} \, t \right)} \right] \, ,
\end{equation}
\begin{equation}\label{eq: fluctuations_psi_eq}
    \psi{\left( \mathbf{r} \right)} = \psi_0 \left[ 1 + u_{\mathbf{k}} \, e^{i \left( \mathbf{k} \cdot \mathbf{r} - \omega_{\mathbf{k}} \, t \right)} + v^*_{\mathbf{k}} \, e^{-i \left( \mathbf{k} \cdot \mathbf{r} - \omega_{\mathbf{k}} \, t \right)} \right] e^{-i \, \omega \, t} \, .
\end{equation}
Inserting the linearized fields~\eqref{eq: fluctuations_n_eq}-\eqref{eq: fluctuations_psi_eq} into the GPE~\eqref{eq: mf_hc_n_eq}-\eqref{eq: mf_hc_psi_eq}, we obtain the eigenvalue equation
\begin{equation}\label{eq: fluctuations_eigen_eq}
\small
    \begin{pmatrix}
    \left( 1 - 2 \, n_0 \right) J{\left( \mathbf{k} \right)} + z \, \bar{U} \, n_0 + \omega_c & 0 & \left[ 2 \, z \, J + \bar{U}{\left( \mathbf{k} \right)} \right] n_0 \\
    0 & \left( 2 \, n_0 - 1 \right) J{\left( \mathbf{k} \right)} - z \, \bar{U} \, n_0 - \omega_c & -\left[ 2 \, z \, J + \bar{U}{\left( \mathbf{k} \right)} \right] n_0 \\
    \left( 1 - n_0 \right) \left[ z \, J + J{\left( \mathbf{k} \right)} \right] & -\left( 1 - n_0 \right) \left[ z \, J + J{\left( \mathbf{k} \right)} \right] & 0
    \end{pmatrix}
    \begin{pmatrix}
    u_{\mathbf{k}} \\
    v_{\mathbf{k}} \\
    w_{\mathbf{k}}
    \end{pmatrix}
    = \omega_{\mathbf{k}}
    \begin{pmatrix}
    u_{\mathbf{k}} \\
    v_{\mathbf{k}} \\
    w_{\mathbf{k}}
    \end{pmatrix}
    \, ,
\end{equation}
where we have defined $\bar{U}{\left( \mathbf{k} \right)} = 2 \, \bar{U} \sum_{a = 1}^{d} \cos{\left( k_a \right)}$. The eigenvalues $\omega_{\mathbf{k}}$ provide the energy spectrum of the collective modes of the hard-core state, the most relevant of which is the Goldstone branch
\begin{equation}\label{eq: goldstone_eq}
    \omega^{\mathrm{hc}}_{\mathrm{G}}{\left( \mathbf{k} \right)} = \sqrt{\frac{\left( z \, \bar{U} + 2 \, \omega_c \right)^2}{z^2 \left( 2 \, J + \bar{U} \right)^2} \left[ z \, J + \varepsilon{\left( \mathbf{k} \right)} \right]^2 + 2 \left| \psi_0 \right|^2  \left[ 2 \, z \, J + \bar{U}{\left( \mathbf{k} \right)} \right] \left[ z \, J + \varepsilon{\left( \mathbf{k} \right)} \right]} \, ,
\end{equation}
which is an acoustic excitation with sound velocity~\cite{PhysRevLett.103.230403}
\begin{equation}\label{hc_goldstone_cs}
    c^{\mathrm{hc}}_s = \sqrt{2 \, z \, J \left( 2 \, J + \bar{U} \right)} \left| \psi_0 \right| = \sqrt{2 \, z \, J \left( 2 \, J + \bar{U} \right) n_0 \left( 1 - n_0 \right)} = \sqrt{\frac{2 \, J \left( z \, J - \omega_c \right) \left[ z \left( J + \bar{U} \right) + \omega_c \right]}{z \left( 2 \, J + \bar{U} \right)}} \, .
\end{equation}
To conclude, we notice that the sound velocity becomes imaginary (signalling a dynamical instability of the ground state) for $\omega_c < -z \left( J + \bar{U} \right)$ and $\omega_c > z \, J$.


\section{Complete expression and properties \texorpdfstring{\\}{} of the Bogoliubov-de Gennes superoperator \texorpdfstring{$\hat{\mathcal{L}}_{\mathbf{k}}$}{}}\label{supsec: L_k}
Fluctuations on top of the NESS are formally expressed as
\begin{equation}\label{eq: GW_fluctuations}
    \vec{c}{\left( \mathbf{r}, t \right)} = \vec{c}_0{\left( t \right)} + \delta \vec{c}{\left( \mathbf{r}, t \right)} = \hat{U}{\left( t \right)} \left[ \vec{c}_0 + \vec{u}_{\mathbf{k}} \, e^{i \left( \mathbf{k} \cdot \mathbf{r} - \omega_{\mathbf{k}} \, t \right)} + \vec{v}^*_{\mathbf{k}} \, e^{-i \left( \mathbf{k} \cdot \mathbf{r} - \omega^*_{\mathbf{k}} \, t \right)} \right] \hat{U}^{\dagger}{\left( t \right)} \, ,
\end{equation}
where the unitary operator $\hat{U}{\left( t \right)} = e^{-i \left( \hat{n} + \hat{\sigma}_z/2 \right) \omega_0 \, t}$ rotates the density matrix in the reference frame of limit cycles (if present). Hence, the matrix elements of the upper diagonal block of the Bogoliubov-de Gennes Lindbladian $\hat{\mathcal{L}}_{\mathbf{k}}$ are given by
\begin{equation}\label{eq: A_block}
\begin{aligned}
    A^{n', m', \mu, \mu'}_{n, m, \sigma, \sigma'}{\left( \mathbf{k} \right)} = &\ L^{n', m', \mu, \mu'}_{n, m, \sigma, \sigma'}{\left[ \vec{c}_0 \right]} + J{\left( \mathbf{k} \right)} \left[ \left( \sqrt{n} \left( c_0 \right)_{n - 1, m, \sigma, \sigma'} - \sqrt{m + 1} \left( c_0 \right)_{n, m + 1, \sigma, \sigma'} \right) \sqrt{n'} \, \delta_{n' - 1, m'} \, \delta_{\mu, \mu'} \right. \\
    &\left. + \left( \sqrt{n + 1} \left( c_0 \right)_{n + 1, m, \sigma, \sigma'} - \sqrt{m} \left( c_0 \right)_{n, m - 1, \sigma, \sigma'} \right) \sqrt{n' + 1} \, \delta_{n' + 1, m'} \, \delta_{\mu, \mu'} \right] \\
    &- \omega_0 \left( n - m + \frac{\sigma - \sigma'}{2} \right) \delta_{n, n'} \, \delta_{m, m'} \, \delta_{\sigma, \mu} \, \delta_{\sigma', \mu'} \, .
\end{aligned}
\end{equation}
The Bogoliubov equations governing the particle $\left( \vec{u}_{\mathbf{k}} \right)$ and hole $\left( \vec{v}_{\mathbf{k}} \right)$ sectors are essentially related by complex conjugation, as a direct consequence of the Hermitian connection between the elements of the density matrix $c_{n, m, \sigma, \sigma'}{\left( \mathbf{r} \right)}$. As a result, we obtain that particle and hole fluctuations are related to each other by transposition through $u_{\alpha, \mathbf{k}, n, m, \sigma, \sigma'} = v_{\alpha, \mathbf{k}, m, n, \sigma', \sigma}$. We additionally notice that, as a byproduct of the global U(1) invariance of the ansatz~\eqref{MAIN-eq: GW_density_matrix}, there is a single zero-energy eigenmode $\omega_{\alpha = 0, \mathbf{k}} = 0$ corresponding to the NESS itself~\cite{castin}. Also, the eigenvalues of $\hat{\mathcal{L}}_{\mathbf{k}}$ with $\text{Re}{\big( \omega_{\alpha, \mathbf{k}} \big)} \neq 0$ can be grouped into anti-conjugate pairs as $\omega_{\alpha, \mathbf{k}} = \pm \omega'_{\alpha, \mathbf{k}} + i \, \omega''_{\alpha, \mathbf{k}}$, since $\text{Re}{\big[ \text{Tr}{\big( \hat{A}_{\mathbf{k}} \big)} \big]} = 0$; however, a less intuitive relation holds between positive- and negative-energy eigenvectors.

For the sake of clarity, we specify that in the present work we never fix the normalization of the two (distinct) components of the right eigenvectors $\left( \vec{u}_{\alpha, \mathbf{k}}, \vec{v}_{\alpha, \mathbf{k}} \right)$, as the semi-analytical expression of the observables of interest will be shown to be always independent of its choice, see Subsection~\ref{supsec: one_body_correlations}(A).


\section{Fluctuations of relevant observables}\label{supsec: fluctuations_observables}
In this Section, we describe in detail how the fluctuation amplitudes of the collective modes in different observables channels are properly defined within our linear response approach.

Inserting the density matrix expansion~\eqref{MAIN-eq: GW_fluctuations} into the expression of Eq.~\eqref{eq: n_formula}, we find that linear fluctuations of the local photon density due to the $\left( \alpha, \mathbf{k} \right)$ excitation behave as
\begin{equation}\label{eq: n_expansion}
    n{\left( \mathbf{r} \right)} = \mathlarger\sum_n \mathlarger\sum_{\sigma} n \left[ \left( c_0 \right)_{n, n, \sigma, \sigma} + u_{\alpha, \mathbf{k}, n, n, \sigma, \sigma} \, e^{i \, \mathbf{k} \cdot \mathbf{r}} + v^*_{\alpha, \mathbf{k}, n, n, \sigma, \sigma} \, e^{-i \, \mathbf{k} \cdot \mathbf{r}} \right] = n_0 + \left( N_{\alpha, \mathbf{k}} \, e^{i \, \mathbf{k} \cdot \mathbf{r}} + \text{c.c.} \right) \, ,
\end{equation}
where we have the \textit{density} spectral weight
\begin{equation}\label{eq: N_definition}
    N_{\alpha, \mathbf{k}} \equiv \sum_n \sum_{\sigma} n \, u_{\alpha, \mathbf{k}, n, n, \sigma, \sigma} \, .
\end{equation}
Similarly, the order parameter is perturbed according to
\begin{equation}\label{eq: psi_expansion}
    \psi{\left( \mathbf{r} \right)} = \mathlarger\sum_n \mathlarger\sum_{\sigma} \sqrt{n + 1} \left[ \left( c_0 \right)_{n + 1, n, \sigma, \sigma} + u_{\alpha, \mathbf{k}, n + 1, n, \sigma, \sigma} \, e^{i \, \mathbf{k} \cdot \mathbf{r}} + v^*_{\alpha, \mathbf{k}, n + 1, n, \sigma, \sigma} \, e^{-i \, \mathbf{k} \cdot \mathbf{r}} \right] = \psi_0 + U_{\alpha, \mathbf{k}} \, e^{i \, \mathbf{k} \cdot \mathbf{r}} + V^*_{\alpha, \mathbf{k}} \, e^{-i \, \mathbf{k} \cdot \mathbf{r}} \, ,
\end{equation}
where
\begin{equation}\label{eq: U}
    U_{\alpha, \mathbf{k}} \equiv \sum_n \sum_{\sigma} \sqrt{n + 1} \, u_{\alpha, \mathbf{k}, n + 1, n, \sigma, \sigma}
\end{equation}
and
\begin{equation}\label{eq: V}
    V_{\alpha, \mathbf{k}} \equiv \sum_n \sum_{\sigma} \sqrt{n + 1} \, v_{\alpha, \mathbf{k}, n + 1, n, \sigma, \sigma} = \sum_n \sum_{\sigma} \sqrt{n + 1} \, u_{\alpha, \mathbf{k}, n, n + 1, \sigma, \sigma}
\end{equation}
measure the \textit{particle} and \textit{hole} character of a given excitation with respect to the photon field, respectively.

We highlight that, whereas particle and hole fluctuations of the density matrix are related by simple transposition $\vec{v}_{\mathbf{k}} = \left( \vec{u}_{\mathbf{k}} \right)^T$, there is no obvious particle-hole symmetry condition for excitations of the photon field, since $\left| U_{\alpha, \mathbf{k}} \right| \neq \left| V_{\alpha, \mathbf{k}} \right|$ in principle. This is due to the fact that $u_{\alpha, \mathbf{k}, n, m, \sigma, \sigma'} \neq u^*_{\alpha, \mathbf{k}, m, n, \sigma', \sigma}$ in general, meaning that either particle or hole fluctuations of the density matrix are not individually bound to be Hermitian. For completeness, we also stress that the spectral amplitudes determined above should be interpreted as \textit{relative} weights of the excitation modes in each excitation channel, since we never specify the normalization of the Bogoliubov components $\left( \vec{u}_{\alpha, \mathbf{k}}, \vec{v}_{\alpha, \mathbf{k}} \right)$.

Using Eq.~\eqref{eq: psi_expansion}, we can also estimate the spectral contribution of each excitation mode to amplitude and phase perturbations of the order parameter. To lowest order, upon writing the order parameter as $\psi{\left( \mathbf{r} \right)} = \left| \psi{\left( \mathbf{r} \right)} \right| \exp{\left[ i \, \varphi{\left( \mathbf{r} \right)} \right]}$, the former kind of fluctuations reads
\begin{equation}\label{eq: psi_amplitude_expansion}
    \delta \left| \psi{\left( \mathbf{r} \right)} \right| \approx \delta \left\{ \left| \psi{\left( \mathbf{r} \right)} \right| \cos{\left[ \varphi{\left( \mathbf{r} \right)} \right]} \right\} = \frac{1}{2} \, \delta \left[ \psi{\left( \mathbf{r} \right)} + \text{c.c.} \right] = \frac{1}{2} \left( U_{\alpha, \mathbf{k}} + V_{\alpha, \mathbf{k}} \right) e^{i \, \mathbf{k} \cdot \mathbf{r}} + \text{c.c.} \, ,
\end{equation}
while phase fluctuations are approximately captured by
\begin{equation}\label{eq: psi_phase_expansion}
    \delta \varphi{\left( \mathbf{r} \right)} \approx \frac{\delta \left\{ \left| \psi{\left( \mathbf{r} \right)} \right| \sin{\left[ \varphi{\left( \mathbf{r} \right)} \right]} \right\}}{\left| \psi{\left( \mathbf{r} \right)} \right|} = \frac{1}{2 \, i \left| \psi{\left( \mathbf{r} \right)} \right|} \delta \left[ \psi{\left( \mathbf{r} \right)} - \text{c.c.} \right] \propto \frac{1}{2 \, i} \left( U_{\alpha, \mathbf{k}} - V_{\alpha, \mathbf{k}} \right) e^{i \, \mathbf{k} \cdot \mathbf{r}} + \text{c.c.}
\end{equation}


\begin{figure}[!t]
    \centering
    \includegraphics[width=0.7\linewidth]{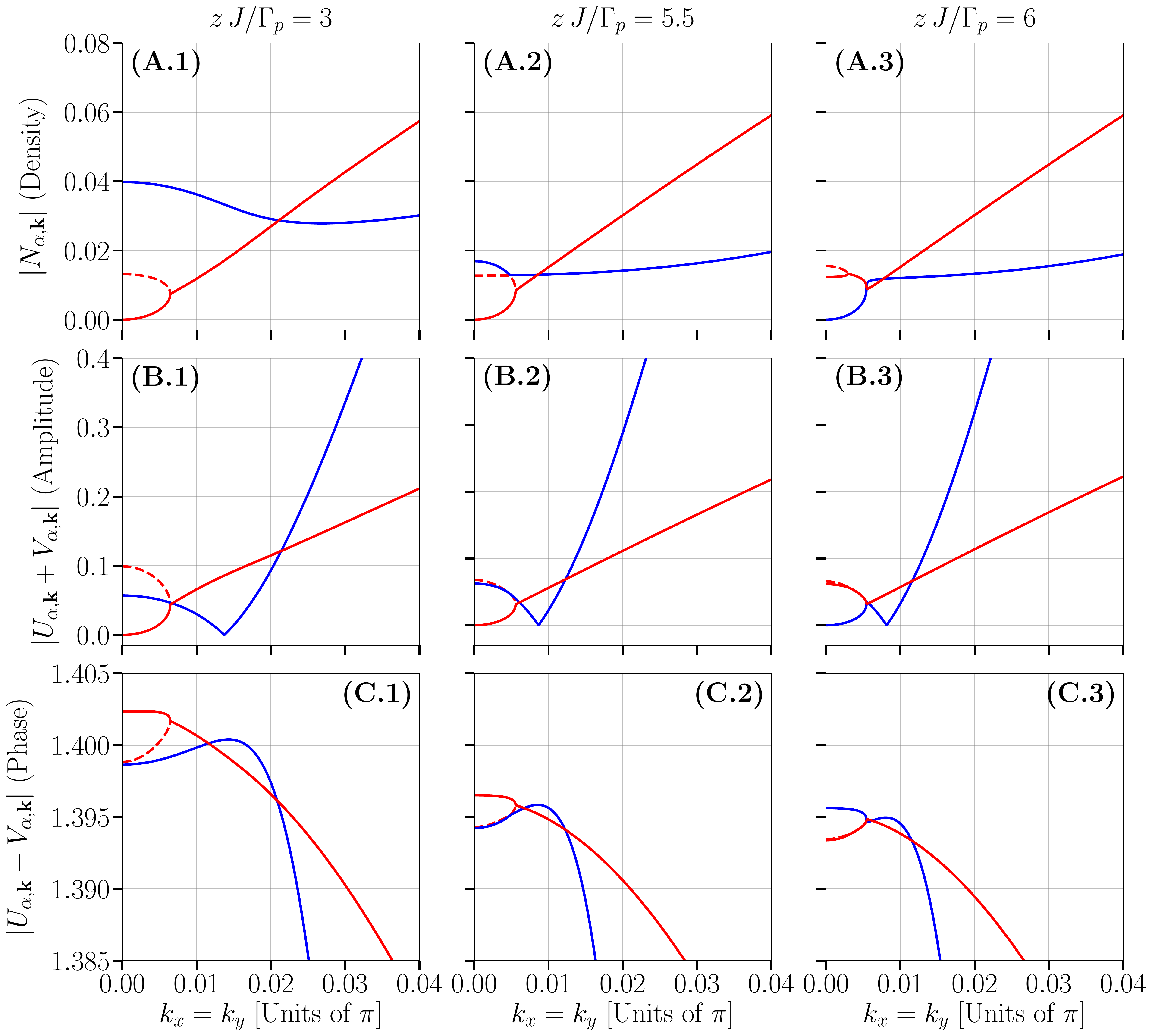}
    \caption{Fluctuation amplitudes of the collective modes of the SFP with respect to different excitation channels for the same parameters of Fig.~\ref{MAIN-fig: spectra} in the main text. Columns from left to right correspond to increasing values of the hopping $J$, in particular: (a)-(a$'$) $z \, J/\Gamma_p = 3$; (b)-(b$'$) $z \, J/\Gamma_p = 5.5 \approx z \, J_m/\Gamma_p$; (c)-(c$'$) $z \, J/\Gamma_p = 10$. Rows from top to bottom: modulus of the density weight $N_{\alpha, \mathbf{k}}$ and amplitude/phase weights $U_{\alpha, \mathbf{k}} \pm V_{\alpha, \mathbf{k}}$ amplitudes for each collective mode.}
    \label{fig: figure_4b}
\end{figure}

\section{Addenda on the excitation spectrum in the superfluid phase}\label{supsec: sf_excitations}
\subsection{Fluctuation amplitudes of the collective modes in the SFP}\label{supsubsec: SFP_fluctuation_amplitudes}
In Fig.~\ref{fig: figure_4b}, we illustrate how the fluctuation amplitudes of the collective modes of the SFP change by increasing $J$ across the anti-adiabatic crossover, which takes place at $z \, J/\Gamma_p \approx 5.5$ for the chosen parameters. Interestingly, we notice that the Goldstone hybridization takes place also at the level of the fluctuation amplitudes, see in particular the comparison between panels (A.2)-(A.3) and (B.2)-(B.3). More precisely, upon increasing the hopping $J$, the D-mode gradually loses its density character and acquires the same amplitude character of the ghost mode in the diffusive momentum region just before the hybridization point. Moreover, in panels (C.1)-(C.3), we explicitly show that the D-mode is always characterized by a large phase character, comparable to the one of the non-hybridized Goldstone branch, and indeed becomes the excitation with the highest phase character after the anti-adiabatic crossover.

As mentioned in the main text, the observed drastic change in the density character of the D-mode across the SFP could indicate that more general dynamical observables probing two-body correlations of the SFP -- corresponding to intensity correlations in the present quantum optical context -- could be the ideal targets of measurement protocols aimed at detecting the spectral properties of the hybridized Goldstone mode. Since the Gutzwiller approximation is known to underestimate pair correlations at the level of Gaussian fluctuations, such a topic goes beyond the scope of the present work, and we leave it as an open problem for future investigations.

\subsection{Additional comments on the anti-adiabatic regime of the SFP}
It is important to highlight that the TLE losses ($\gamma$) turn out to have little effect on the spectral properties discussed previously: increasing $\gamma/\Gamma_p$ has the simple outcome of spoiling population inversion in the IP and pushing the critical hopping $J_c$ to larger values. It is instructive to compare this result with what happens in an exciton-polariton condensate when dissipation of the reservoir polaritons is faster than losses in the condensate: in that context, the Goldstone mode strongly entangles with the dissipative channels as well, but the condensate is prone to density modulations due to the repulsive interaction with the reservoir~\cite{PhysRevLett.99.140402, Baboux:18}. By contrast, in the present case phase excitations are always stable in the anti-adiabatic regime, independently of the ratio $\gamma/\Gamma_l$. This suggests a different physical origin for the present Goldstone hybridization: indeed, we argue that the Rabi coupling with the TLE's is key to a stable mixing of dissipation and coherent dynamics at large $J$. As a side note, we point out that such anti-adiabatic limit of the SFP does not admit a simple Gross-Pitaevskii description. This would in fact require a mean-field decoupling of the Rabi interaction and hence would neglect all those local fluctuations that are instead fully incorporated by our Gutzwiller theory.

\begin{figure}[!t]
    \centering
    \includegraphics[width=0.8\linewidth]{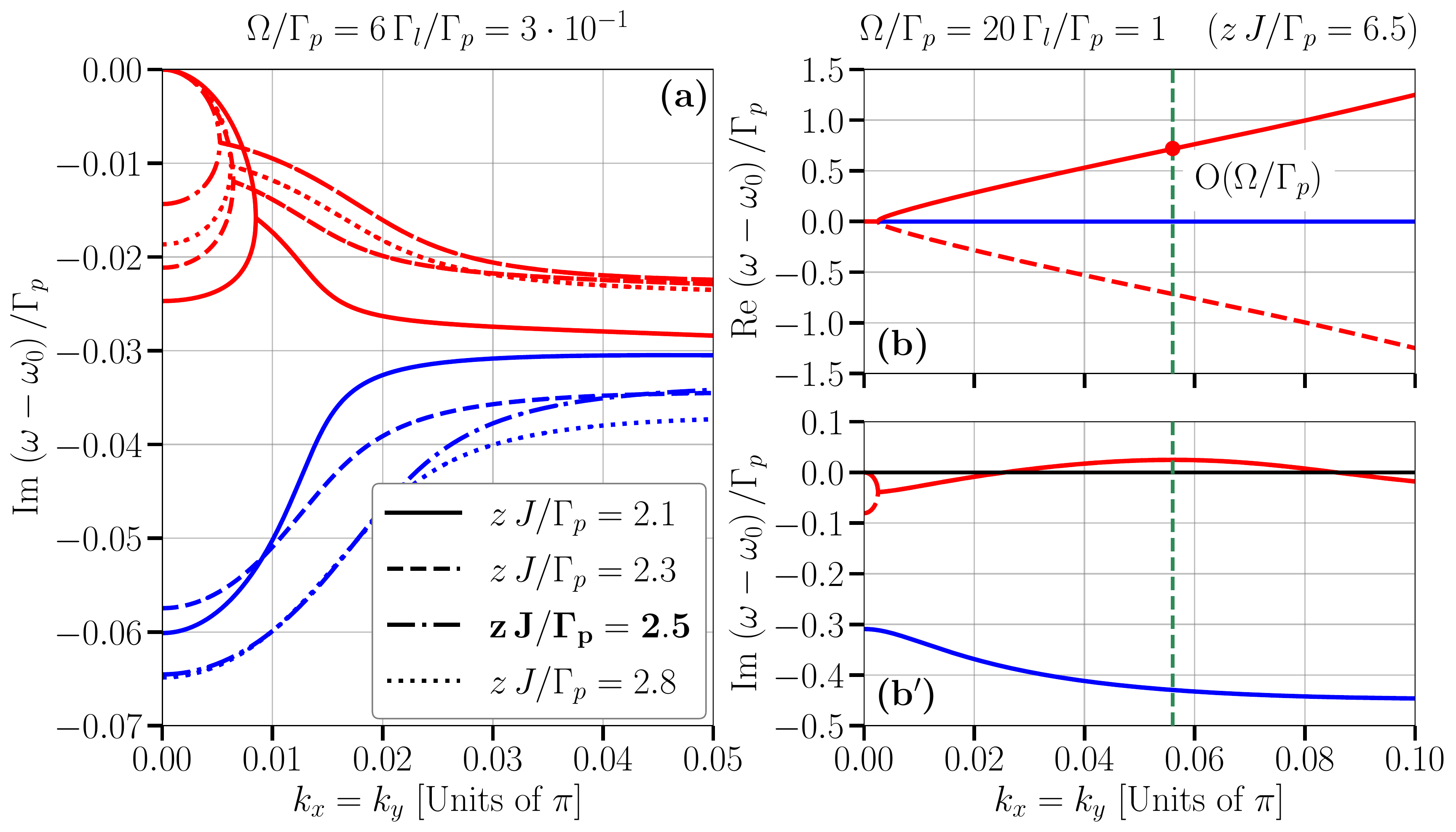}
    \caption{Panel (a): imaginary part of the excitation spectrum of the SF state for $\Omega/\Gamma_p = 5 \cdot 10^{-1}$, $\Gamma_l/\Gamma_p = 5 \cdot 10^{-2}$ and $\gamma/\Gamma_p = \cdot 10^{-3}$. Different line styles refer to distinct hopping values in a neighbourhood of the absolute maximum of the condensate density $\rho_c = \left| \psi_0 \right|^2$ appearing in the strong-coupling regime, see panel (b) of Fig.~\ref{fig: figure_1_SM}. The hopping value $z \, J/\Gamma_p \approx 2.5$ corresponds to the maximum of $\rho_c$ and the largest energy separation between the amplitude mode damping $\Gamma_{\mathrm{A}}{\left( \mathbf{k} \right)}$ and the D-mode damping $\Gamma_{\mathrm{D}}{\left( \mathbf{k} \right)}$.
    Panels (b)-(b$'$): excitation spectrum of the SF state stabilized for $\Omega/\Gamma_p = 1$, $\Gamma_l/\Gamma_p = 5 \cdot 10^{-2}$ and $\gamma/\Gamma_p = \cdot 10^{-3}$ with the hopping energy $z \, J/\Gamma_p = 6.5$, inside the equivalent of the cyan-shaded area in Fig.~\ref{fig: figure_1_SM}(b). The dashed green line indicates the incommensurate momentum for which the NESS is maximally unstable. As in the main text, the expression $\omega - \omega_0$ indicates that the excitation energy is calculated with respect to the lasing frequency $\omega_0$.
    }
    \label{fig: figure_3_SM}
\end{figure}

\subsection{Peculiarities of the SFP spectrum in the strong-coupling regime}\label{supsubsec: SFP_spectrum_strong_coupling}
In panel (a) of Fig.~\ref{fig: figure_3_SM}, we illustrate how the developing of an absolute maximum in the condensate density in the strong-coupling regime (associated with a purification of the NESS into the true hard-core state at equilibrium) reflects into the excitation spectrum, focusing in particular on the damping rates of the collectives modes. Starting from values of $J$ just beyond the critical point ($z \, J/\Gamma_p = 2.1$), we observe that the energy scale of the amplitude damping decreases significantly and reaches its minimal value exactly where the maximum of the condensate density is located ($z \, J/\Gamma_p = 2.5$). At the same time, the D-mode damping reaches larger values. Thus, the maximal separation between the two branches takes place at the optimal hopping. It must be noted that also the radius of the sphere of diffusive momenta displays the same kind of behavior.

Whereas in the previous Subsection we have shown that the SFP is not affected by dynamical instabilities in the dilute regime of the system, a different scenario is obtained when considering the superfluid states preceding the anti-adiabatic crossover $\left( J \lesssim J_m \right)$ in the strong-coupling limit, as illustrated in panels (b) and (b$'$). Here, we find that the Goldstone damping rate $\Gamma_{\mathrm{G}}{\left( \mathbf{k} \right)}$ becomes positive over a finite momentum range if the effective TLE coupling $G = \Omega^2/\left( \Gamma_p \, \Gamma_l \right)$ exceeds a specific threshold. Additional proof of the prominent physical role played by the light-matter interaction is given by the fact that the Goldstone energy $\varepsilon_{\mathrm{G}}{\left( \mathbf{k} \right)}$ corresponding to the maximum value of $\Gamma_{\mathrm{G}}{\left( \mathbf{k} \right)}$ [see the red point in panel (b) and the green dashed line in panel (b$'$)] scales linearly with $\Omega$.


\section{Calculation of linear response functions}\label{supsec: linear_response}
\subsection{General equations}\label{supsubsec: general_response}
Let us consider a generic perturbation in the momentum-frequency channel $\left( \mathbf{k}, \omega \right)$ represented by the operator $\hat{F}_{\mathbf{r}}$,
\begin{equation}\label{eq: perturbation}
    \hat{H}_{\mathrm{P}}{\left( t \right)} \equiv \mathlarger\sum_{\mathbf{r}} \left[ V_{\mathbf{k}, \omega} \, e^{i \left( \mathbf{k} \cdot \mathbf{r} - \omega \, t \right)} \hat{F}_{\mathbf{r}} + V^*_{\mathbf{k}, \omega} \, e^{-i \left( \mathbf{k} \cdot \mathbf{r} - \omega \, t \right)} \hat{F}^{\dagger}_{\mathbf{r}} \right] \, .
\end{equation}
Gaussian fluctuations induced by such perturbation can be evaluated by considering its contribution at linear order to the dynamical equations of quantum fluctuations~\eqref{MAIN-eq: GW_eigen}, which simply generalize into
\begin{equation}\label{eq: lr_eq_1}
    \omega
    \begin{pmatrix}
    \vec{u}_{\mathbf{k}} \\
    \vec{v}_{\mathbf{k}}
    \end{pmatrix}
    = \hat{\mathcal{L}}_{\mathbf{k}}
    \begin{pmatrix}
    \vec{u}_{\mathbf{k}} \\
    \vec{v}_{\mathbf{k}}
    \end{pmatrix}
    +
    \hat{W}_{\mathbf{k}, \omega}
    \begin{pmatrix}
    \vec{c}_0 \\
    \left( \vec{c}_0 \right)^*
    \end{pmatrix}
    \, ,
\end{equation}
where $\hat{W}_{\mathbf{k}, \omega} \equiv \text{diag}{\big( \hat{w}_{1, \mathbf{k}}, -\hat{w}^*_{2, \mathbf{k}} \big)}$, with the matrix blocks $\hat{w}_{1, \mathbf{k}} \sim V_{\mathbf{k}, \omega}$ and $\hat{w}_{2, \mathbf{k}} \sim V^*_{\mathbf{k}, \omega}$ given by the Lindbladian terms corresponding the perturbation operators $\hat{F}_{\mathbf{r}}$ and $\hat{F}^{\dagger}_{\mathbf{r}}$ respectively. By inverting Eq.~\eqref{eq: lr_eq_1}, we obtain a convoluted extension of the original Bogoliubov-de Gennes eigenvalue problem,
\begin{equation}\label{eq: lr_eq_2}
    \begin{pmatrix}
    \vec{u}_{\mathbf{k}} \\
    \vec{v}_{\mathbf{k}}
    \end{pmatrix}
    =
    \left( \omega \, \hat{\mathds{1}} - \hat{\mathcal{L}}_{\mathbf{k}} \right)^{-1}
    \begin{pmatrix}
    \hat{w}_{1, \mathbf{k}, \omega} & 0 \\
    0 & -\hat{w}_{2, \mathbf{k}, \omega}
    \end{pmatrix}
    \begin{pmatrix}
    \vec{c}_0 \\
    \left( \vec{c}_0 \right)^*
    \end{pmatrix}
    \, .
\end{equation}
To unfold the previous expression, we resort to the standard spectral theory of Lindbladian operators~\cite{Arrigoni2018, secli_thesis} to conveniently introduce the left eigenvectors $\left( \vec{x}_{\alpha, \mathbf{k}}, \vec{y}_{\alpha, \mathbf{k}} \right)$ of the Lindbladian $\hat{\mathcal{L}}_{\mathbf{k}}$ as
\begin{equation}
    \begin{pmatrix}
    \vec{x}_{\alpha, \mathbf{k}} \\
    \vec{y}_{\alpha, \mathbf{k}}
    \end{pmatrix}
    ^{\dagger}
    \hat{\mathcal{L}}_{\mathbf{k}} = \omega_{\alpha, \mathbf{k}} \begin{pmatrix}
    \vec{x}_{\alpha, \mathbf{k}} \\
    \vec{y}_{\alpha, \mathbf{k}}^{\dagger}
    \end{pmatrix} \, ,
\end{equation}
which form an orthonormal basis and a complete set with the right eigenvectors $\left( \vec{u}_{\alpha, \mathbf{k}}, \vec{v}_{\alpha, \mathbf{k}} \right)$, such that
\begin{equation}
    \begin{pmatrix}
    \vec{x}_{\alpha, \mathbf{k}} \\
    \vec{y}_{\alpha, \mathbf{k}}
    \end{pmatrix}
    ^{\dagger}
    \begin{pmatrix}
    \vec{u}_{\beta, \mathbf{k}} \\
    \vec{v}_{\beta, \mathbf{k}}
    \end{pmatrix}
    = \delta_{\alpha, \beta}
    \quad , \quad
    \mathlarger\sum_{\alpha}
    \begin{pmatrix}
    \vec{x}_{\alpha, \mathbf{k}} \\
    \vec{y}_{\alpha, \mathbf{k}}
    \end{pmatrix}
    ^{\dagger}
    \begin{pmatrix}
    \vec{u}_{\alpha, \mathbf{k}} \\
    \vec{v}_{\alpha, \mathbf{k}}
    \end{pmatrix}
    = \hat{\mathds{1}} \, .
\end{equation}
Therefore, making use of the spectral decomposition of $\hat{\mathcal{L}}_{\mathbf{k}}$
\begin{equation}
    \hat{\mathcal{L}}_{\mathbf{k}} = \mathlarger\sum_{\alpha}
    \begin{pmatrix}
    \vec{u}_{\alpha, \mathbf{k}} \\
    \vec{v}_{\alpha, \mathbf{k}}
    \end{pmatrix}
    \omega_{\alpha, \mathbf{k}}
    \begin{pmatrix}
    \vec{x}_{\alpha, \mathbf{k}} \\
    \vec{y}_{\alpha, \mathbf{k}}
    \end{pmatrix}
    ^{\dagger} \, ,
\end{equation}
we can rewrite Eq.~\eqref{eq: lr_eq_2} as
\begin{equation}\label{eq: lr_eq_3}
\begin{aligned}
    \begin{pmatrix}
    \vec{u}_{\mathbf{k}} \\
    \vec{v}_{\mathbf{k}}
    \end{pmatrix}
    &= \mathlarger\sum_{\alpha}
    \begin{pmatrix}
    \vec{u}_{\alpha, \mathbf{k}} \\
    \vec{v}_{\alpha, \mathbf{k}}
    \end{pmatrix}
    \frac{1}{\omega - \omega_{\alpha, \mathbf{k}}}
    \begin{pmatrix}
    \vec{x}_{\alpha, \mathbf{k}} \\
    \vec{y}_{\alpha, \mathbf{k}}
    \end{pmatrix}
    ^{\dagger}
    \begin{pmatrix}
    \hat{w}_{1, \mathbf{k}, \omega} & 0 \\
    0 & -\hat{w}_{2, \mathbf{k}, \omega}
    \end{pmatrix}
    \begin{pmatrix}
    \vec{c}_0 \\
    \left( \vec{c}_0 \right)^*
    \end{pmatrix}
    \\
    &= \hat{R} \cdot \hat{\Omega}{\left( \omega \right)} \cdot \left[ \hat{R}^{-1} \cdot \hat{W}_{\mathbf{k}, \omega} \cdot
    \begin{pmatrix}
    \vec{c}_0 \\
    \left( \vec{c}_0 \right)^*
    \end{pmatrix}
    \right] \, ,
\end{aligned}
\end{equation}
where $\hat{\Omega}_{\alpha \beta}{\left( \omega \right)} = \delta_{\alpha, \beta}/\left( \omega - \omega_{\alpha, \mathbf{k}} \right)$ and the matrix $\hat{R}$ gathers the right eigenvectors of $\hat{\mathcal{L}}_{\mathbf{k}}$ on its columns. After carrying out the calculation of the right-hand side of Eq.~\eqref{eq: lr_eq_3}, the linear response function for a given observable $\hat{O}{\left( \mathbf{r}, t \right)}$ is directly provided by the corresponding linear expansion in terms of the fluctuation amplitudes $\left( \underline{u}_{\mathbf{k}}, \underline{v}_{\mathbf{k}} \right)$, see Section~\ref{supsec: fluctuations_observables}.

It is important to observe that Eq.~\eqref{eq: lr_eq_3} is independent of the choice of the normalization of the right/left eigenvectors of $\hat{\mathcal{L}}_{\mathbf{k}}$. In fact, fixing the normalization of the two components of the right eigenvectors to some number $\mathcal{N}_{\mathbf{k}}$, the left eigenvectors have to scale with $\mathcal{N}^{-1}_{\mathbf{k}}$, since the latter are given by the rows of $\hat{R}^{-1}$.

\subsection{Density fluctuations from the Bragg response}\label{supsubsec: bragg_response}
As a first illustrative example, let us consider a Bragg perturbation
\begin{equation}\label{eq: bragg_perturbation}
    \hat{H}_{\mathrm{Bragg}}{\left( t \right)} \equiv \sum_{\mathbf{r}} V_{\mathbf{k}, \omega} \, \cos{\left( \mathbf{k} \cdot \mathbf{r} - \omega \, t \right)} \, \hat{n}_{\mathbf{r}} \, .
\end{equation}
This Hamiltonian perturbation produces a modulation in the photon density according to the identity $\delta \langle \hat{n}_{\mathbf{r}} \rangle = \rho_{\mathbf{k}, \omega} \, e^{i \left( \mathbf{k} \cdot \mathbf{r} - \omega \, t \right)} + \text{c.c.}$, where $\rho_{\mathbf{k}, \omega} = \chi_n{\left( \mathbf{k}, \omega \right)} \, V_{\mathbf{k}, \omega}$ with $\chi_n{\left( \mathbf{k}, \omega \right)}$ being defined as the density response function. In this case, Eq.~\eqref{eq: lr_eq_2} specializes into
\begin{equation}\label{eq: lr_eq_n_1}
    \begin{pmatrix}
    \vec{u}_{\mathbf{k}} \\
    \vec{v}_{\mathbf{k}}
    \end{pmatrix}
    = \frac{1}{2} \mathlarger\sum_{\alpha}
    \begin{pmatrix}
    \vec{u}_{\alpha, \mathbf{k}} \\
    \vec{v}_{\alpha, \mathbf{k}}
    \end{pmatrix}
    \frac{1}{\omega - \omega_{\alpha, \mathbf{k}}}
    \begin{pmatrix}
    \vec{x}_{\alpha, \mathbf{k}} \\
    \vec{y}_{\alpha, \mathbf{k}}
    \end{pmatrix}
    ^{\dagger}
    \begin{bmatrix}
    \vec{\mathcal{N}}_0 \\
    -\left( \vec{\mathcal{N}}_0 \right)^*
    \end{bmatrix}
    V_{\mathbf{k}, \omega} \, ,
\end{equation}
where $\left( \mathcal{N}_0 \right)_{n, m, \sigma, \sigma'} = \left( n - m \right) \left( c_0 \right)_{n, m, \sigma, \sigma'}$. The density fluctuation amplitude $\rho_{\mathbf{k}, \omega} \equiv N_{\mathbf{k}}$ is given by contracting the left-hand side of Eq.~\eqref{eq: lr_eq_n_1} by a tensor with elements $n \, \delta_{n, n'} \, \delta_{m, m'} \, \delta_{n, m} \, \delta_{\sigma, \mu} \, \delta_{\sigma', \mu'} \, \delta_{\sigma, \sigma'}$ in both the $\vec{u}_{\mathbf{k}}$ and $\vec{v}_{\mathbf{k}}$ sectors, hence
\begin{equation}\label{eq: lr_eq_n_2}
\begin{aligned}
    2 \, \rho_{\mathbf{k}, \omega} &= \mathlarger\sum_{\alpha}
    \frac{N_{\alpha, \mathbf{k}}}{\omega - \omega_{\alpha, \mathbf{k}}}
    \begin{pmatrix}
    \vec{x}_{\alpha, \mathbf{k}} \\
    \vec{y}_{\alpha, \mathbf{k}}
    \end{pmatrix}
    ^{\dagger}
    \begin{bmatrix}
    \vec{\mathcal{N}}_9 \\
    -\left( \vec{\mathcal{N}}_0 \right)^*
    \end{bmatrix}
    V_{\mathbf{k}, \omega} \\
    &= \mathlarger\sum_{\alpha}
    \frac{N_{\alpha, \mathbf{k}} \left[ \vec{x}^*_{\alpha, \mathbf{k}} \cdot \vec{\mathcal{N}}_0 - \vec{y}^*_{\alpha, \mathbf{k}} \cdot \left( \vec{\mathcal{N}}_0 \right)^* \right]}{\omega - \omega_{\alpha, \mathbf{k}}} \, V_{\mathbf{k}, \omega} \\
    &= 2 \underbrace{\mathlarger\sum_{\alpha}
    \frac{N_{\alpha, \mathbf{k}} \left( \vec{x}^*_{\alpha, \mathbf{k}} \cdot \vec{\mathcal{N}}_0 \right)}{\omega - \omega_{\alpha, \mathbf{k}}}
    }_{\chi_n{\left( \mathbf{k}, \omega \right)}} \, V_{\mathbf{k}, \omega} \, .
\end{aligned}
\end{equation}
Therefore, we observe that the density spectral weight $N_{\alpha, \mathbf{k}}$ sets directly the strength of the dynamical density response of the system in the NESS. We remark that, since the analytic continuation of $\chi_n{\left( \mathbf{k}, \omega \right)}$ can be identified with the two-particle Green's function of the system, a suitable manipulation of its Fourier transform provides the lowest-order estimation of spatial and temporal density correlations.


\section{Linear response of one-body operators}\label{supsec: one_body_correlations}
\subsection{Calculation of the retarded Green's function}\label{supsubsec: retarded_G}
Similarly to the case of a density perturbation, we start our derivation by studying the response of the photon field to a perturbation creating a particle (removing a hole) with a given momentum in the NESS, namely
\begin{equation}\label{eq: ph_perturbation}
    \hat{H}_{\mathrm{p}}{\left( t \right)} \equiv \frac{1}{2} \mathlarger\sum_{\mathbf{r}} \left[ \eta_{\mathbf{k}, \omega} \, e^{i \left( \mathbf{k} \cdot \mathbf{r} - \omega \, t \right)} \, \hat{a}^{\dagger}_{\mathbf{r}} + \eta^*_{\mathbf{k}, \omega} \, e^{-i \left( \mathbf{k} \cdot \mathbf{r} - \omega \, t \right)} \, \hat{a}_{\mathbf{r}} \right] \, ,
\end{equation}
which breaks the U(1) symmetry of the model explicitly and therefore is coupled to fluctuations of the order parameter $\psi{\left( \mathbf{r} \right)}$. The linear response equations corresponding to the perturbation~\eqref{eq: ph_perturbation} have the form
\begin{equation}\label{eq: lr_eq_psi_1}
    \begin{pmatrix}
    \vec{u}_{\mathbf{k}} \\
    \vec{v}_{\mathbf{k}}
    \end{pmatrix}
    = \frac{1}{2} \mathlarger\sum_{\alpha}
    \begin{pmatrix}
    \vec{u}_{\alpha, \mathbf{k}} \\
    \vec{v}_{\alpha, \mathbf{k}}
    \end{pmatrix}
    \frac{1}{\omega - \omega_{\alpha, \mathbf{k}}}
    \begin{pmatrix}
    \vec{x}_{\alpha, \mathbf{k}} \\
    \vec{y}_{\alpha, \mathbf{k}}
    \end{pmatrix}
    ^{\dagger}
    \begin{bmatrix}
    \vec{\mathcal{P}}_0 \\
    -\left( \vec{\mathcal{Q}}_0 \right)^*
    \end{bmatrix}
    \eta_{\mathbf{k}, \omega} \, ,
\end{equation}
where we have defined the vectorized matrices 
\begin{equation}
    \left( \mathcal{P}_0 \right)_{n, m, \sigma, \sigma'} = \sqrt{n} \left( c_0 \right)_{n - 1, m, \sigma, \sigma'} - \sqrt{m + 1} \left( c_0 \right)_{n, m + 1, \sigma, \sigma'}
\end{equation}
and
\begin{equation}
    \left( \mathcal{Q}_0 \right)_{n, m, \sigma, \sigma'} = \sqrt{n + 1} \left( c_0 \right)_{n + 1, m, \sigma, \sigma'} - \sqrt{m} \left( c_0 \right)_{n, m - 1, \sigma, \sigma'} \, .
\end{equation}
Now, the previous expression allows us to calculate two different types of response functions, either in the (i) particle or in the (ii) hole channel.

The first kind of response requires to determine the \textit{particle} fluctuation amplitude $U_{\alpha, \mathbf{k}}$ defined in Eq.~\eqref{eq: U}: this is given by contracting both the sides of Eq.~\eqref{eq: lr_eq_psi_1} by an operator with tensor components $\sqrt{n + 1} \, \delta_{n + 1, n'} \, \delta_{n, m'} \, \delta_{n, m + 1} \, \delta_{\sigma, \mu} \, \delta_{\sigma', \mu'} \, \delta_{\sigma, \sigma'}$ in the $\vec{u}_{\mathbf{k}}$ sector and $\sqrt{n + 1} \, \delta_{n, n'} \, \delta_{n + 1, m'} \, \delta_{n, m - 1} \, \delta_{\sigma, \mu} \, \delta_{\sigma', \mu'} \, \delta_{\sigma, \sigma'}$ in the $\vec{v}_{\mathbf{k}}$ sector. As a result, we obtain
\begin{equation}\label{eq: lr_eq_psi_2}
\begin{aligned}
    2 \, \Psi_{\mathbf{k}, \omega}
    &= \mathlarger\sum_{\alpha} \frac{U_{\alpha, \mathbf{k}}}{\omega - \omega_{\alpha, \mathbf{k}}}
    \begin{pmatrix}
    \vec{x}_{\alpha, \mathbf{k}} \\
    \vec{y}_{\alpha, \mathbf{k}}
    \end{pmatrix}
    ^{\dagger}
    \begin{bmatrix}
    \vec{\mathcal{P}}_0 \\
    -\left( \vec{\mathcal{Q}}_0 \right)^*
    \end{bmatrix}
    \eta_{\mathbf{k}, \omega} \\
    &= \mathlarger\sum_{\alpha}
    \frac{U_{\alpha, \mathbf{k}} \left[ \vec{x}^*_{\alpha, \mathbf{k}} \cdot \vec{\mathcal{P}}_0 - \vec{y}^*_{\alpha, \mathbf{k}} \cdot \left( \vec{\mathcal{Q}}_0 \right)^* \right]}{\omega - \omega_{\alpha, \mathbf{k}}} = \\
    &= 2 \underbrace{\mathlarger\sum_{\alpha}
    \frac{U_{\alpha, \mathbf{k}} \left( \vec{x}^*_{\alpha, \mathbf{k}} \cdot \vec{\mathcal{P}}^0 \right)}{\omega - \omega_{\alpha, \mathbf{k}}}
    }_{G_R{\left( \mathbf{k}, \omega \right)}} \, \eta_{\mathbf{k}, \omega} \, ,
\end{aligned}
\end{equation}
where $\Psi_{\mathbf{k}, \omega}$ is the order parameter variation. Physically speaking, the response of the order parameter to the perturbation~\eqref{eq: ph_perturbation} in the particle channel can be interpreted as the normal component of the retarded Green's function of cavity photons~\cite{krutitsky_navez, Stringari2018}. More explicitly, the explicit expression of our prediction for the Green's function is
\begin{equation}\label{eq: G_R}
    G_R{\left( \mathbf{k}, \omega \right)} \equiv \mathlarger\sum_{\alpha}
    \frac{Z_{\alpha, \mathbf{k}}}{\omega - \omega_{\alpha, \mathbf{k}}} = \mathlarger\sum'_{\alpha}
    \left[ \frac{Z_{\alpha, \mathbf{k}}}{\omega - \omega_{\alpha, \mathbf{k}}} + \frac{Y^*_{\alpha, \mathbf{k}}}{\omega + \omega^*_{\alpha, \mathbf{k}}} \right] \, ,
\end{equation}
which has been written in a more symmetric form in the last equality. Here, we have defined the quasiparticle
\begin{equation}
    Z_{\alpha, \mathbf{k}} \equiv U_{\alpha, \mathbf{k}} \left( \vec{x}^*_{\alpha, \mathbf{k}} \cdot \vec{\mathcal{P}}_0 \right)
\end{equation}
and quasihole
\begin{equation}
    Y_{\alpha, \mathbf{k}} \equiv V_{\alpha, \mathbf{k}} \left( \vec{y}^*_{\alpha, \mathbf{k}} \cdot \vec{\mathcal{P}}^*_0 \right)
\end{equation}
weights, while the summation on the right-hand side of Eq.~\eqref{eq: G_R} is restricted to excitations with positive real energy.

The second type of dynamical fluctuations which can be drawn out of Eq.~\eqref{eq: lr_eq_psi_2} encodes the response of the order parameter in the hole channel. This corresponds to extracting the \textit{hole} amplitude $V_{\alpha, \mathbf{k}}$ from the right-hand side of Eq.~\eqref{eq: lr_eq_psi_1} in the same way as outlined above for the normal component. The final result of this procedure is the retarded anomalous component of the Green's function, having the expression
\begin{equation}\label{eq: F_R}
    \Delta_R{\left( \mathbf{k}, \omega \right)} \equiv \mathlarger\sum_{\alpha}
    \frac{\overline{Z}_{\alpha, \mathbf{k}}}{\omega - \omega_{\alpha, \mathbf{k}}} = \mathlarger\sum'_{\alpha}
    \left[ \frac{\overline{Z}_{\alpha, \mathbf{k}}}{\omega - \omega_{\alpha, \mathbf{k}}} + \frac{\overline{Y}^*_{\alpha, \mathbf{k}}}{\omega + \omega^*_{\alpha, \mathbf{k}}} \right] \, ,
\end{equation}
where we have introduced the anomalous quasiparticle $\overline{Z}_{\alpha, \mathbf{k}} = \left( V_{\alpha, \mathbf{k}}/U_{\alpha, \mathbf{k}} \right) Z_{\alpha, \mathbf{k}}$ and quasihole $\overline{Y}_{\alpha, \mathbf{k}} = \left( U_{\alpha, \mathbf{k}}/V_{\alpha, \mathbf{k}} \right) Y_{\alpha, \mathbf{k}}$ weights. As one could expect by physical intuition, anomalous correlations play a major role in the SF phase of the NESS and, in analogy with the case of exciton-polariton condensates~\cite{PhysRevB.79.125311}, can be exploited for directly probing the excitation spectrum of the system, as we will discuss more in depth in the following Subsections~\ref{supsec: one_body_correlations}(C)-(D).

Applying the concepts of Keldysh field theory~\cite{Sieberer_2016}, the simplest object provided by the retarded Green's function is the DoS of the NESS, reading
\begin{equation}\label{eq: DOS}
    A{\left( \mathbf{k}, \omega \right)} \equiv -\frac{1}{\pi} \, \text{Im}{\left[ G_R{\left( \mathbf{k}, \omega \right)} \right]} = -\frac{1}{\pi} \mathlarger\sum_{\alpha} \frac{\omega''_{\alpha, \mathbf{k}} \, Z'_{\alpha, \mathbf{k}} + \left( \omega - \omega'_{\alpha, \mathbf{k}} \right)  Z''_{\alpha, \mathbf{k}}}{\left( \omega - \omega'_{\alpha, \mathbf{k}} \right)^2 + \left( \omega''_{\alpha, \mathbf{k}} \right)^2}
\end{equation}
where the symbols $'$ and $''$ indicate real and imaginary parts respectively. It is worth noticing that the DoS is not a plain weighted sum of Lorentz distributions $g{\left( \omega \right)} \propto \Lambda/\big[ \left( \omega - \omega_p \right)^2 + \Lambda^2 \big]$ centered around the real parts of the poles of the Green's function (as it would happen at equilibrium), but in principle could get a finite contribution from terms with the functional form $f{\left( \omega \right)} \propto \left( \omega - \omega_p \right)/\big[ \left( \omega - \omega_p \right)^2 + \Lambda^2 \big]$, which is an odd function of $\omega$ with respect to $\omega'_{\alpha, \mathbf{k}}$ and therefore corresponds to a sort of Fano resonance of the collective modes. This is indeed the case of the DoS profile shown in \autoref{MAIN-fig: T_R}(b) of the main text and is a crucial consequence of the fact that the quasiparticle weight $Z_{\alpha, \mathbf{k}}$ can generically acquire a complex value out of equilibrium, since it quantifies no longer the overlap of a collective mode with a single-particle excitation of the NESS, but depends on the spectral decomposition of the Lindbladian in a non-trivial way~\cite{scarlatella2019}. In particular, this is inherently connected to the fact that right and left eigenvectors are not related by simple conjugation, meaning that the creation and destruction of an elementary excitation on top of the stationary state are not inverse processes. This is the reason why the sign of $A{\left( \mathbf{k}, \omega \right)}$ can become negative when the NESS is far from being an equilibrium configuration of the system, which we have shown to be the case of the IP in the main text. In this respect, we notice that the local DoS $A{\left( \omega \right)} \equiv V^{-1} \sum_{\mathbf{k}} A{\left( \mathbf{k}, \omega \right)}$ satisfies the sum rule
\begin{equation}\label{eq: DOS_sum_rule}
    \int d\omega \, A{\left( \omega \right)} = 1 - 2 \left\langle \hat{n} \right\rangle
\end{equation}
as a product of the commutation relation between hard-core bosonic operators. This automatically implies that $A{\left( \mathbf{k}, \omega \right)} < 0$ for some values of $\left( \mathbf{k}, \omega \right)$ at least for $n_0 > 1/2$: once again, this hints at the fundamental role of strong interactions in building the insulating regime of the model in combination with strong dissipation~\cite{PhysRevX.11.031018}.

\subsection{Dynamical response to a weak probe: transmittivity, reflectivity and four-wave mixing functions}\label{supsubsec: cavity_response}
In this Subsection, we briefly review the input-output theory of optical cavities in the specific context of our driven-dissipative setting in order to introduce the response functions of interest in our study, namely those giving the transmittivity and reflectivity of the cavity array.

Along the same conceptual lines of Subsection~\ref{supsec: one_body_correlations}(A), an alternative quantum description of the driving of a cavity photon mode by an incident coherent light beam can be obtained by resorting to the so-called \textit{input-output theory}~\cite{gardiner_zoller, walls_milburn} for optical cavities~\cite{PhysRevA.74.033811}. In particular, the Hamiltonian term describing the external driving of a standard two-sided cavity by an incident field of amplitude $E_{\mathrm{in}}{\left( \mathbf{r}, t \right)}$ is akin to the perturbation of Eq.~\eqref{eq: ph_perturbation} and can be written in $\mathbf{k}$-space as
\begin{equation}
    \hat{H}_{\mathrm{drive}}{\left( t \right)} \equiv i \mathlarger\sum_{\mathbf{k}} \left[ \eta_{\mathrm{F}, \mathbf{k}} \, \tilde{E}_{\mathrm{in}}{\left( \mathbf{k}, t \right)} \, \hat{a}^{\dagger}_{\mathbf{k}} - \eta^*_{\mathrm{F}, \mathbf{k}} \, \tilde{E}^*_{\mathrm{in}}{\left( \mathbf{k}, t \right)} \, \hat{a}_{\mathbf{k}} \right] \, ,
\end{equation}
where $\tilde{E}_{\mathrm{in}}{\left( \mathbf{k}, t \right)}$ is the Fourier transform of $E_{\mathrm{in}}{\left( \mathbf{r}, t \right)}$ and $\eta_{\mathrm{F}, \mathbf{k}}$ is the transmission amplitude of the front mirror of the cavity. In the following, we will also denote the transmission amplitude of the back mirror by $\eta_{\mathrm{B}, \mathbf{k}}$. We remark that these coefficients are physically linked with the radiative damping $\Gamma_l$ by the simple relation $2 \, \Gamma_l = \left| \eta_{\mathrm{F}, \mathbf{k}} \right|^2 + \left| \eta_{\mathrm{B}, \mathbf{k}} \right|^2$. For the sake of simplicity, we always assume that the cavity has a fully symmetric geometry such that the transmission amplitudes read $\left| \eta_{\mathrm{F/B}, \mathbf{k}} \right| = \sqrt{\Gamma_l}$ and are independent of momentum. We also recall here that $\eta_{\mathrm{F/B}, \mathbf{k}}$ can be usually extracted from transmission and reflection measurements on the unloaded cavity.

The finite transmittivity of the front and the back mirrors of the cavity is responsible for the re-emission of light with an amplitude proportional to the intra-cavity field $\left\langle \hat{a}_{\mathbf{k}} \right\rangle$. By means of the boundary conditions set by the two cavity mirrors, in the linear-response regime the reflected and transmitted fields can be related to the intra-cavity field within the input-output framework as~\cite{RevModPhys.85.299, chiocchetta2017}
\begin{subequations}
\begin{equation}
    \tilde{E}_{\mathrm{T}}{\left( \mathbf{k}, \omega \right)} = T{\left( \mathbf{k}, \omega \right)} \, \tilde{E}_{\mathrm{in}}{\left( \mathbf{k}, \omega \right)}
\end{equation}
\begin{equation}
    \tilde{E}_{\mathrm{R}}{\left( \mathbf{k}, \omega \right)} = R{\left( \mathbf{k}, \omega \right)} \, \tilde{E}_{\mathrm{in}}{\left( \mathbf{k}, \omega \right)}
\end{equation}
\end{subequations}
respectively, where we have now formally introduced the transmission
\begin{equation}
    T{\left( \mathbf{k}, \omega \right)} \equiv -i \, \eta_{\mathrm{F}, \mathbf{k}} \, \eta^*_{\mathrm{B}, \mathbf{k}} \, G_R{\left( \mathbf{k}, \omega \right)}
\end{equation}
and reflection
\begin{equation}
    R{\left( \mathbf{k}, \omega \right)} \equiv 1 - i \left| \eta_{\mathrm{F}, \mathbf{k}} \right|^2 G_R{\left( \mathbf{k}, \omega \right)} = 1 + \left( \frac{\eta_{\mathrm{F}, \mathbf{k}}}{\eta_{\mathrm{B}, \mathbf{k}}} \right)^* T{\left( \mathbf{k}, \omega \right)}
\end{equation}
functions discussed in the main text. Thus, we can write the expression of the reflectivity as
\begin{equation}\label{R_explicit_expression}
    \left| R{\left( \mathbf{k}, \omega \right)} \right|^2 = \left[ 1 - \pi \left| \eta_{\mathrm{F}, \mathbf{k}} \right|^2 A{\left( \mathbf{k}, \omega \right)} \right]^2 + \left| \eta_{\mathrm{F}, \mathbf{k}} \right|^4 \text{Re}{\left[ G_R{\left( \mathbf{k}, \omega \right)} \right]}^2 \, ,
\end{equation}
which recovers the result of Eq.~\eqref{MAIN-eq: R} in the main text.

Importantly, it must be noted that the standard sum rule
\begin{equation}\label{T_R_sum_rule}
    \left| T{\left( \mathbf{k}, \omega \right)} \right|^2 + \left| R{\left( \mathbf{k}, \omega \right)} \right|^2 = 1 + 2 \left| \eta_{\mathrm{F}, \mathbf{k}} \right|^2 \left\{ \Gamma_l \left| G_R{\left( \mathbf{k}, \omega \right)} \right|^2 + \text{Im}{\left[ G_R{\left( \mathbf{k}, \omega \right)} \right]} \right\} \stackrel{?}{=} 1
\end{equation}
is fulfilled only if $\Gamma_l \left| G_R{\left( \mathbf{k}, \omega \right)} \right|^2 = -\text{Im}{\left[ G_R{\left( \mathbf{k}, \omega \right)} \right]} = \pi \, A{\left( \mathbf{k}, \omega \right)}$. Whereas this condition is usually satisfied in many common situations at equilibrium, it could be instead largely violated in the presence of peculiar out-of-equilibrium effects, for instance when the NESS is characterized by spontaneous energy emission. In the main text and in Subsection~\eqref{supsec: one_body_correlations}(B), we have shown that this is not only the case of the SFP, where the $\mathbf{k = 0}$ lasing state naturally enhances both $T{\left( \mathbf{k}, \omega \right)}$ and $R{\left( \mathbf{k}, \omega \right)}$, but remarkably also of the IP, for which we uncover an anomalous behavior of the response functions as a consequence of the population inversion phenomenon. Indeed, since we know that $A{\left( \mathbf{k}, \omega \right)} < 0$ for some specific values of $\left( \mathbf{k}, \omega \right)$ across the whole IP, we straightforwardly obtain that $\left| R{\left( \mathbf{k}, \omega \right)} \right|^2 > 1$ from Eq.~\eqref{R_explicit_expression} and that Eq.~\eqref{T_R_sum_rule} is always violated in this regime, as given by the numerical results presented in Fig.~\ref{MAIN-fig: T_R} of the main text.

\begin{figure}[!t]
    \centering
    \includegraphics[width=0.45\linewidth]{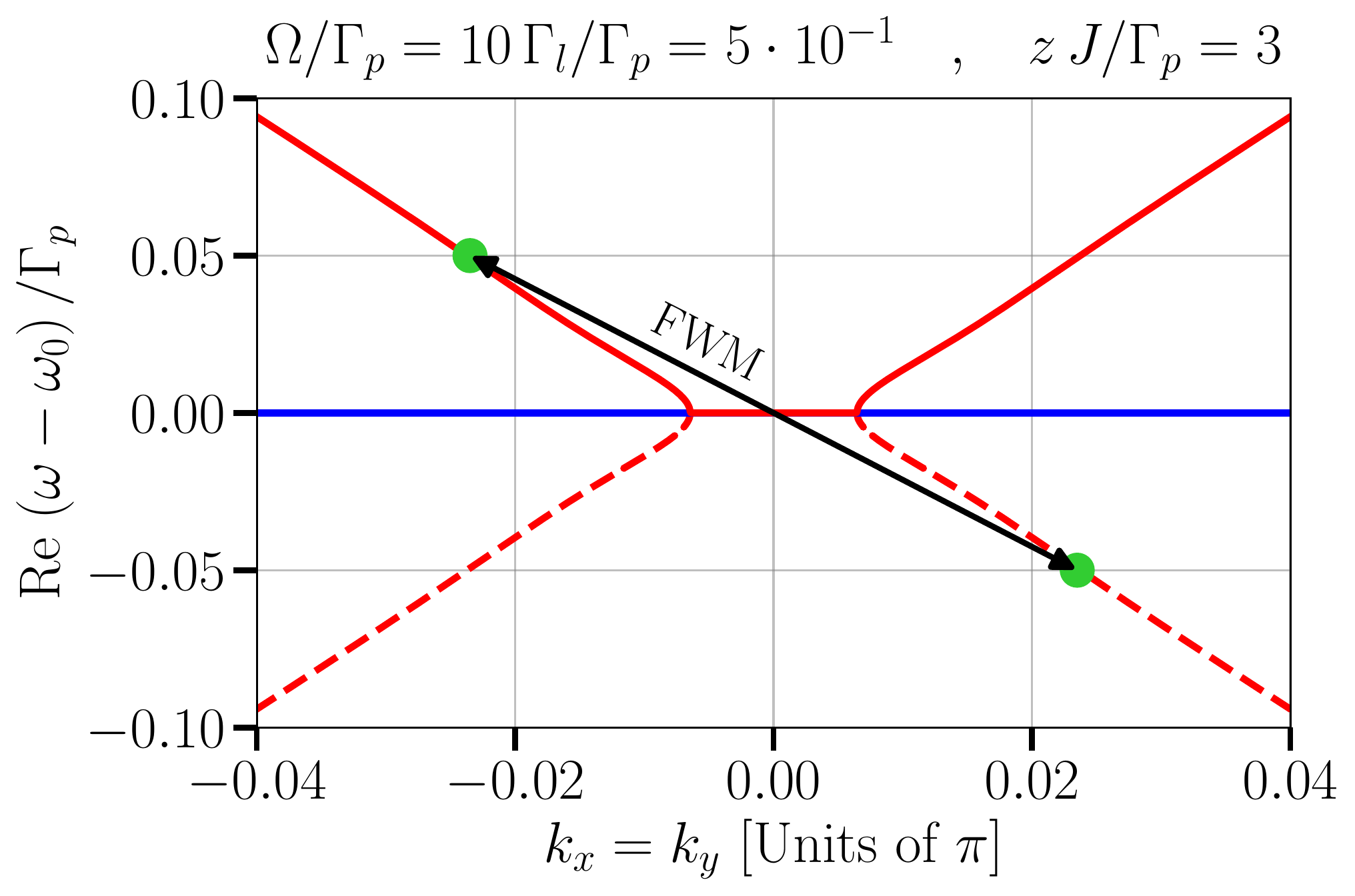}
    \caption{Pictorial sketch of the FWM measurement protocol. The colored lines represent an instance of the collective excitations of the SFP. The black arrows illustrate the scattering process from which the FWM signal is generated: the system is perturbed at $\left( \mathbf{k}, \omega \right)$ (lower green dot) and its response is probed at $\left( -\mathbf{k}, -\omega \right)$.}
    \label{fig: figure_fwm}
\end{figure}

We conclude this Subsection by introducing a third useful response function of interest in the SFP, going under the name of Four-Wave Mixing (FWM) signal, which has been proven useful to attain a solid experimental evidence of the amplitude excitation branch in exciton-polariton condensates~\cite{PhysRevB.79.125311}. The physical process underlying the measurement protocol of the FWM response is sketched in Fig.~\ref{fig: figure_fwm}. Elementary excitations are created on top of the condensate by injecting extra photons with a probe laser beam with a finite momentum $\mathbf{k}$ and tuned at a frequency $\omega$. The response of the system is then observed via the coherent light emission at an opposite wave vector $-\mathbf{k}$ and energy $2 \, \omega_0 - \omega$: the existence of a coherent coupling between the frequencies $\omega$ and $2 \, \omega_0 - \omega$ (located symmetrically around the effective chemical potential $\omega_c$) and the momenta $\pm \mathbf{k}$ from the fact that the elementary
excitations of the condensate consist of a coherent superposition of plane waves at $\left( \mathbf{k}, \omega \right)$ and $\left( -\mathbf{k}, 2 \, \omega_0 - \omega \right)$: this can be in turn interpreted as a clue of the existence of anomalous correlations in the system. Indeed, it turns out that the FWM signal is simply provided by the retarded anomalous propagator,
\begin{equation}
    F{\left( \mathbf{k}, \omega \right)} \equiv -i \, \eta_{\mathrm{F}, \mathbf{k}} \, \eta^*_{\mathrm{B}, \mathbf{k}} \, \Delta_R{\left( \mathbf{k}, \omega \right)}
\end{equation}
which has been shown to couple positive- and negative-energy modes of the system, see Eq.~\eqref{eq: F_R} and the following Subsection.

\subsection{Comparison between the DoS and the transmittivity spectrum in the IP}\label{supsubsec: T_IP}
\begin{figure}[!t]
    \centering
    \includegraphics[width=0.6\linewidth]{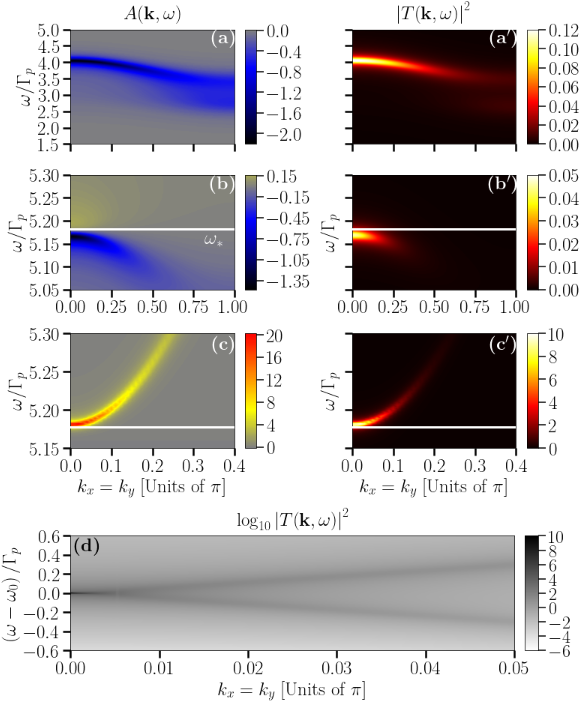}
    \caption{Spectral profiles of the DoS (left column) and the transmittivity (right column) for the same parameters considered in \autoref{MAIN-fig: T_R} of the main text. In particular, $\Omega/\Gamma_p = 5 \cdot 10^{-1}$, $\Gamma_l/\Gamma_p = 5 \cdot 10^{-2}$ and $\gamma/\Gamma_p = 5 \cdot 10^{-3}$. From top to bottom, the hopping energy is (a)-(a$'$) $z \, J/\Gamma_p = 0.5$, (b)-(b$'$) $z \, J/\Gamma_p = 2$ and (c)-(c$'$) $z \, J/\Gamma_p = 2.5$.}
    \label{fig: figure_5a}
\end{figure}

In this Subsection, we provide a more complete discussion of the DoS structure and transmittivity spectra in the IP to integrate the discussion on dynamical response functions in the main text. In particular, in \autoref{fig: figure_5a} we place side by side the DoS and $\left| T{\left( \mathbf{k}, \omega \right)} \right|^2$ profiles as done for the reflectivity.

In the Mott-like regime at low $J$ [Fig.~\ref{fig: figure_5a}(a)], most part of $A{\left( \mathbf{k}, \omega \right)}$ lies well below the effective chemical potential $\omega_*$ and has a dual profile depending on the momentum of QP excitations. In particular, whereas the DoS reaches its minimal value at low momenta, it covers a wider range of states at the border of the Brillouin zone, where it splits into two peaks. This behavior can be elegantly explained in terms of the key role played by the quasiparticle weight $Z_{+, \mathbf{k}}$ in the DoS expression~\eqref{eq: DOS}. The real part of $Z_{+, \mathbf{k}}$, associated with the Lorentzian component of $A{\left( \mathbf{k}, \omega \right)}$, is \textit{negative} and gives a leading contribution for non-local QP states, which explains the well-visible peak of the DoS at small momenta. By contrast, the imaginary part of $Z_{+, \mathbf{k}}$, which instead weighs the \textit{odd} resonance of QP modes, favours localized states and yields the corresponding double-peaked profile of the DoS for $\left| \mathbf{k} \right| \approx \pi$. However, since we always have $|Z'_{+, \mathbf{k}}| > |Z''_{+, \mathbf{k}}|$ at small $J$, the latter contribution is never sufficiently large to flip the DoS sign in the Mott-like regime. Therefore, we realize that both the negativity of $Z'_{+, \mathbf{k}}$ and a significant imaginary component $Z''_{+, \mathbf{k}}$ strongly relate to the appearance of population inversion and, more generally, hint at the strong non-equilibrium character of the NESS. The shape of the DoS carries over to the transmittivity [Fig.~\ref{fig: figure_5a}(a$'$)]. However, the latter function shows no signature of the non-equilibrium effects due to population inversion, in stark contrast with the reflectivity spectrum discussed in the main text.

As $J$ increases, population inversion is progressively lost: this manifests into the QP band nearing the frequency threshold $\omega_*$ starting from non-local states at $\mathbf{k} = \boldsymbol{0}$. In particular, as underlined in the main text, we notice that states lying above $\omega_*$ acquire a significant positive weight. Once again, we can quantitatively understand this mechanism from the point of view of the quasiparticle weight. Moving towards the point where the QP band flattens out and $n_0 = 1/2$ [Fig.~\ref{fig: figure_5a}(b)], both the components of $Z_{+, \mathbf{k}}$ decrease in amplitude towards comparable values and become flat in momentum. Thus, the DoS has precisely the shape of a Fano resonance around the QP energy $\omega_{+, \mathbf{k}} \approx \omega_*$, determined by the imaginary part of $Z_{+, \mathbf{k}}$. As before, the anomalous features of the DoS have little effect on $\left| T{\left( \mathbf{k}, \omega \right)} \right|^2$ [Fig.~\ref{fig: figure_5a}(b$'$)]: the only byproduct of the spectral redistribution of the DoS is the non-monotonic dependence of the transmittivity, which uniformly decreases as compared to its small $J$ values and starts accumulating at $\omega_*$ as expected.

Increasing further $J$ towards the Mott/superfluid transition [Fig.~\ref{fig: figure_5a}(c)], the real part of the quasiparticle weight becomes large and positive for states for which $\omega > \omega_*$, while the imaginary component remains a vanishingly small number and gives the residual negative DoS below $\omega_*$. Ultimately, the DoS becomes strictly non-negative exactly before the critical point $J = J_c$. Here, the whole spectral weight has been transferred above the effective chemical potential $\omega_*$: then, this frequency scale can be rigorously identified with the critical energy of delocalized QP excitations, which are then free to condense. The transmittivity profile quite mimics the very same behavior of the DoS [Fig.~\ref{fig: figure_5a}(c$'$)], reaching values above 1 \textit{already} inside the IP. Indeed, the IP critical state close to the transition is the only regime where \textit{both} the transmittivity and the reflectivity are found to allow for light amplification in the absence of long-range coherence, anticipating the physics of the lasing state.

\subsection{More details on the dynamical response of the SFP}\label{supsubsec: response_SFP}
\begin{figure}[!t]
    \centering
    \includegraphics[width=0.85\linewidth]{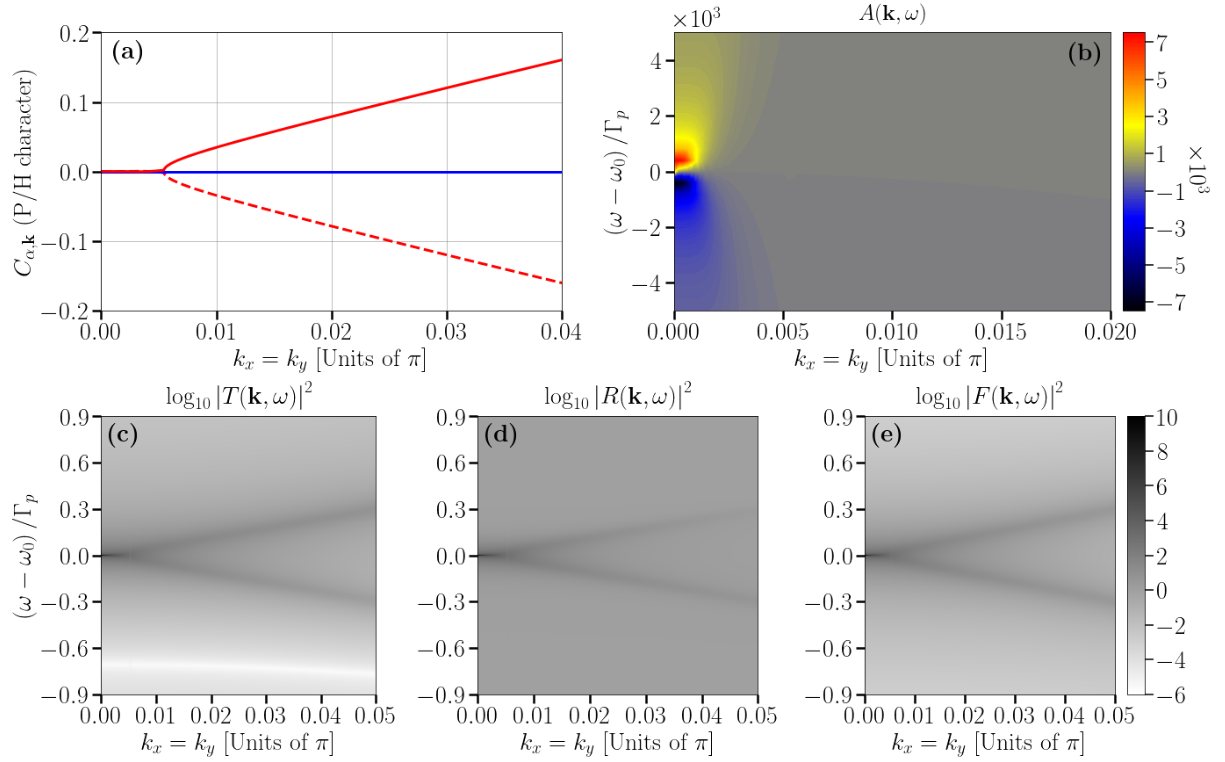}
    \caption{Panel (a): particle-hole character $C_{\alpha, \mathbf{k}} = \left( \left| U_{\alpha, \mathbf{k}} \right| - \left| V_{\alpha, \mathbf{k}} \right| \right)/\left( \left| U_{\alpha, \mathbf{k}} \right| + \left| V_{\alpha, \mathbf{k}} \right| \right)$ of the collective modes of the SFP. Specifically, $C_{\alpha, \mathbf{k}} = \pm 1$ for pure particle (hole) excitations, respectively.
    Panel (b): typical profile of the DoS in the SFP.
    Panels (c)-(e): behavior of the transmittivity, reflectivity and FWM signals in the SFP, respectively.
    All the panels refer to the representative case of the SF state at $z \, J/\Gamma_p = 10$ for $\Omega/\Gamma_p = 5 \cdot 10^{-1}$, $\Gamma_l/\Gamma_p = 5 \cdot 10^{-2}$ and $\gamma/\Gamma_p = 10^{-3}$, see the excitation spectrum in Fig.~\ref{MAIN-fig: spectra}(b)-(b$'$) of the main text.}
    \label{fig: figure_7}
\end{figure}

We start our supplementary analysis of dynamical response in the SFP by looking at the DoS. In Fig.~\ref{fig: figure_7}~(b), we report the typical form of $A{\left( \mathbf{k}, \omega \right)}$ in the symmetry-broken phase, including the anti-adiabatic limit of the model. On the whole, the DoS gets a non-negligible contribution from the Goldstone branch only and exhibits a butterfly shape with a perfectly asymmetric structure around $\omega = \omega_0$. On the one hand, states at $\omega > 0$ have a non-vanishing and positive distribution in the range of diffusive momenta $\Delta \mathbf{k}_{\mathrm{diff}}$, which is peaked around $\left( \mathbf{k = 0}, \omega = 0 \right)$ because of condensation, such that it connects continuously to the DoS of quasiparticle excitations of the critical IP [see Fig.~\ref{fig: figure_5a}~(d)]; on the other hand, states belonging to the ghost branch at $\omega < 0$ have a negative weight which mirrors exactly $A{\left( \mathbf{k}, \omega > 0 \right)}$.

Such a reflection symmetry is another indirect consequence of the particle-hole symmetry characterizing the Goldstone and amplitude excitations at low momenta. This feature is explicitly shown in Fig.~\ref{fig: figure_7}~(a), where we plot the \textit{particle-hole character} $\mathcal{C}_{\alpha, \mathbf{k}} \equiv \left( \left| U_{\alpha, \mathbf{k}} \right| - \left| V_{\alpha, \mathbf{k}} \right| \right)/\left( \left| U_{\alpha, \mathbf{k}} \right| + \left| V_{\alpha, \mathbf{k}} \right| \right)$ of the SFP excitations. In particular, we highlight that the particle/hole amplitudes of the excitation modes are all equal and satisfy $\left| U_{\alpha, \mathbf{k}} \right| = \left| V_{\alpha, \mathbf{k}} \right|$ in the window of diffusive momenta: this means that the non-equilibrium superfluid state, in spite of its non-trivial particle-hole character depending on $J$, is characterized by an emergent particle-hole symmetry on long-range spatial scales. Such a symmetry is then gradually broken by the Goldstone and ghost branches only as the momentum is increased, with the two excitations acquiring predominantly a particle and a hole character, respectively.

For the sake of clarity, we remark that the odd behavior of the DoS around $\omega = 0$ is a genuine product of the Goldstone diffusivity out of equilibrium and must not be confused with the well-known DoS structure of positive/negative-norm modes in an interacting bosonic system at equilibrium: in particular, we find that an essential ingredient for the functional form of the DoS in the SFP is again the imaginary component of the quasiparticle weight of the Goldstone mode $Z''_{G, \mathbf{k}} < 0$, which is negligible everywhere but in the region of diffusive momenta as expected. Hence, we find that the DoS is well-approximated by the expression
\begin{equation}
    A{\left( \mathbf{k}, \omega \right)} = -\frac{1}{\pi} \frac{Z''_{G, \mathbf{k}} \, \omega}{\omega^2 + \left( \omega''_{G, \mathbf{k}} \right)^2} \, ,
\end{equation}
from which we can extract also a prediction for the local DoS at low frequency by analytical integration. In $d$ dimensions, we obtain
\begin{equation}
    A{\left( \omega \right)} = -\frac{1}{\pi} \frac{\Omega_d}{\left( 2 \, \pi \right)^d} \frac{Z''_{\mathrm{G}}}{\sqrt{D}} \arctan{\left( \frac{\sqrt{D} \, \Lambda_{\mathrm{D}}}{\omega} \right)} \, ,
\end{equation}
where we have assumed $Z''_{G, \mathbf{k}} \approx Z''_{\mathrm{G}}$, $D$ is the Goldstone diffusion coefficient and $\Lambda_{\mathrm{D}}$ is a momentum cutoff enclosing the diffusive momentum regime where $\omega''_{G, \mathbf{k}} \approx -D \, \mathbf{k}^2$. Notably, the static limit of the local DoS reads $A{\left( \omega \to 0 \right)} \sim Z''_{\mathrm{G}}/\sqrt{D}$ and provides direct information on the diffusion coefficient. We remind here that $\omega''_{G, \mathbf{k}}$ and $Z''_{G, \mathbf{k}}$ refer also to the hybridized D-mode becoming gapless in the anti-adiabatic regime of the SFP.

As already illustrated in the main text, the very same symmetry that governs the DoS behavior at low energy reflects in the response functions of the photon field, whose profiles are compared in Fig.~\ref{fig: figure_7}(c)-(e) for the same value of $J$ as the DoS. Differently from the case of exciton-polariton condensates, the signals of the upper and lower branches have equally strong intensities in both the transmittivity and the reflectivity spectra, with a weak asymmetry appearing when the particle-hole symmetry of the excitations is increasingly broken at large momenta. Thus, the inherent symmetry properties of the hard-core NESS make the $T{\left( \mathbf{k}, \omega \right)}$, $R{\left( \mathbf{k}, \omega \right)}$ and $F{\left( \mathbf{k}, \omega \right)}$ as equally sensitive probes of the low-energy excitations of the superfluid state of the photon fluid. A secondary property of the transmittivity spectrum is the appearance of a dark resonance for $\omega < 0$ [see Fig.~\ref{fig: figure_7}(c)], well below the signal of the ghost mode, which moves to larger energies and acquires a broader dispersion as $J$ increases. For this reason, we interpret this feature as the effect of the destructive interference between the amplitude mode and the Goldstone/D-mode branches at the level of the Green's function, postponing a deeper understanding of such finer aspects to a more detailed study of the SFP dynamical behavior.

\bibliographystyle{apsrev4-1}
\bibliography{refs}

\makeatletter\@input{work_file_SM.tex}\makeatother